РОССИЙСКАЯ АКАДЕМИЯ НАУК
**ВЫЧИСЛИТЕЛЬНЫЙ ЦЕНТР**

---

**А.В. Гасников**

# СРАВНЕНИЕ ОПРЕДЕЛЕНИЙ ОБОБЩЕННОГО РЕШЕНИЯ ЗАДАЧИ КОШИ ДЛЯ КВАЗИЛИНЕЙНОГО УРАВНЕНИЯ

В работе сопоставляются различные определения обобщенного решения начальной задачи Коши для квазилинейного уравнения. Основной результат заключается в том, что существует и притом единственное обобщенное решение, которое удовлетворяет всем приведенным в работе определениям. Этот результат позволяет сопоставлять решения квазилинейного уравнения и его дифференциально-разностного аналога (уравнение Полтеровича–Хенкина), играющего важную роль в математической экономике. Второе издание было подготовлено в 2009 году, в нем были немного обновлены ссылки на литературу. В настоящем издании были уточнены результаты, касающиеся определения обобщенного решения по Шананину–Хенкину–Полтеровичу (конец третьего параграфа).

**Работа выполнена при поддержке**
- **Российского фонда фундаментальных исследований (код проекта 15-31-70001 мол_а_мос);**
- **Российского научного фонда (проект 16-11-10246)**

Рецензенты: Н.С. Петросян,
И.Г. Поспелов

Научное издание

# ВВЕДЕНИЕ

В работе сопоставляются различные определения обобщенного решения задачи Коши для квазилинейного уравнения типа закона сохранения (1), (2)

$$\frac{\partial \eta(u)}{\partial t} + \frac{\partial \varphi(u)}{\partial x} = 0 \ , \tag{1}$$

$$u\big|_{t=0} = u_0(x). \tag{2}$$

Проблема определения решения задачи Коши (1), (2), теоремы существования и единственности, непрерывная зависимость решения от начальных данных, асимптотические свойства решений изучались, в основном, для случая $\eta(u) = u$. Естественно было бы предположить, что рассматриваемый нами случай $\eta'(u) > 0$, с помощью каких-то преобразований сводится к хорошо изученному $\eta(u) = u$. Например,

1) если разделить уравнение (1) на $\eta'(u) \geq a > 0$;

2) сделать замену $v = \eta(u)$.

Однако чтобы обосновать возможность таких преобразований, мы должны сначала определиться с тем, что мы понимаем под решением. Заметим, что задача (1), (2) может не иметь классических решений.

Рассмотрим отдельно предлагаемые преобразования.

1) В физике уравнение (1), как правило, описывает некоторый закон сохранения (условие непрерывности), причем получается он обычно в интегральном виде, а потом, уже из интегрального вида, путем предельного перехода, приходят в точках гладкости решения к виду дифференциальному [стр. 44 – 47, 1]. Понятно, что если мы будем умножать уравнение (1) на произвольную достаточно гладкую функцию, то с точки зрения дифференциальных законов сохранения, мы будем получать эквивалентные уравнения. Однако нам интересны и разрывные решения. Поэтому, умножив (1) на гладкую функцию, мы должны перейти к интегральному закону сохранения, породившему получившийся дифференциальный. Но, как легко убедится, умножая на разные функции, мы будем приходить к разным интегральным законам сохранения и, следовательно, к разным скоростям движения разрыва (формула Гюгонио–Ренкина [2]).

2) Замена $v=\eta(u)$, приведет (1), (2) к хорошо изученному виду

$$\frac{\partial v}{\partial t}+\frac{\partial\varphi\left(\eta^{-1}(v)\right)}{\partial x}=0 \ . \tag{3}$$

$$v\big|_{t=0}=\eta\left(u_0(x)\right). \tag{4}$$

Если под решением задачи Коши (1), (2) понимать предел почти всюду при $\varepsilon\to 0+$ решений $u_\varepsilon(t,x)$ задачи Коши (5), (6)

$$\frac{\partial\eta(u_\varepsilon)}{\partial t}+\frac{\partial\varphi(u_\varepsilon)}{\partial x}=\varepsilon\frac{\partial^2 u_\varepsilon}{\partial x^2} \ , \tag{5}$$

$$u_\varepsilon\big|_{t=0}=u_0(x). \tag{6}$$

то решения задач Коши (1), (2) и (3), (4) будут связаны соотношением $v=\eta(u)$ тогда и только тогда, когда

$$\frac{\partial v_\varepsilon^1}{\partial t}+\frac{\partial\varphi\left(\eta^{-1}\left(v_\varepsilon^1\right)\right)}{\partial x}=\varepsilon\frac{\partial^2\eta^{-1}\left(v_\varepsilon^1\right)}{\partial x^2},\ \frac{\partial v_\varepsilon^2}{\partial t}+\frac{\partial\varphi\left(\eta^{-1}\left(v_\varepsilon^2\right)\right)}{\partial x}=\varepsilon\frac{\partial^2 v_\varepsilon^2}{\partial x^2},$$

$$v_\varepsilon^1(0,x)=v_\varepsilon^2(0,x)=\eta\left(u_0(x)\right),$$

$$\lim_{\varepsilon\to 0+}v_\varepsilon^1(t,x)=\lim_{\varepsilon\to 0+}v_\varepsilon^2(t,x) \text{ – почти всюду.}$$

Таким образом, ответ на вопрос связаны ли решения задач Коши (1), (2) и (3), (4) соотношением $v=\eta(u)$, сводится к решению частного случая задачи, поставленной И.М. Гельфандом в работе [2].

Суть в том, что далеко не любое определение обобщенного решения удобно для исследования замены, предложенной во втором пункте. Оказывается, есть такое удачное определение (С.Н. Кружков дал его в конце 60-ых), обобщив которое со случая $\eta(u)=u$ на случай $\eta'(u)>0$, можно строго показать, что замена $v=\eta(u)$, действительно, сводит изучение задач с $\eta'(u)>0$ к задачам с $\eta(u)=u$. Однако в разных работах используются разные определения, и, чтобы можно было обобщить основные результаты с $\eta(u)=u$ на $\eta'(u)>0$, нужно показать эквивалентность различных определений определению С.Н. Кружкова.

В настоящей работе предпринята попытка, в хронологическом порядке привести (или, хотя бы, отметить) значимые и представляющие исторический интерес определения обобщенного решения задачи Коши (1), (2).[1] Объяснить происхождение различных

---

[1] При этом мы рассматриваем далеко не полный набор, известных к настоящему моменту определений – так, например, в работе не отражены результаты L. Tartar'а по компенсированной компактности, результаты Ph. Bénilan'а, который использовал нелинейную теорию полугрупп; мало внимания уделено кинетическим формулировкам энтропийных решений, а также связи с нелинейными стохастическими уравнениями в частных производных и численными методами. С современным состоянием дел в теории обобщенных решений квазилинейных уравнений типа закона сохранения и систем таких уравнений можно ознакомиться по следующим работам (и цитированной в них литературе):

1. Эванс Л.К. Методы слабой сходимости для нелинейных уравнений с частными производными. Новосибирск, 2006.
2. Lions P.-L. О некоторых интригующих проблемах нелинейных уравнений в частных производных. Математика: границы и перспективы, М.: ФАЗИС, 2005, стр.193-211.
3. Dafermos C.M. Hyperbolic conservation laws in continuum physics. Springer, 2005.
4. Тупчиев В.А. Обобщенные решения законов сохранения. М.: Наука, 2006.
5. Holden H., Risebro N.H. Front tracking for hyperbolic conservation laws. Springer, 2007.
6. Лакс П.Д. Гиперболические дифференциальные уравнения в частных производных. М.-Ижевск: НИЦ "Регулярная и хаотическая динамика", 2010.

Помимо обобщенных решений можно определять и сингулярные решения (допускаются разрывы решений не только первого рода, но и в виде дельта функции, и её производной) квазилинейных уравнений типа закона сохранения и систем таких уравнений – см. Шелкович В.М. Сингулярные решения систем законов сохранения типа $\delta$ и $\delta'$-ударных волн и процессы переноса и концентрации // УМН, т.63, № 3(381), (2008).

определений. Показать, при некоторых условиях, их эквивалентность, посмотреть какими свойствами обладают обобщенные решения. Следует заметить, что большое внимание уделяется раннему периоду исследований (50-ые, 60-ые годы XX века), поскольку именно в этот период были получены основные результаты. Также заметим, что если какое-то определение дается при определенных условиях, например, начальная функция монотонная или $\varphi\left(\eta^{-1}(v)\right)$ – строго выпуклая функция, то эквивалентность этого определения остальным, следует понимать так, что и другие определения даются при тех же условиях.

Ключевую роль при доказательстве эквивалентности определений играет следующее достаточно очевидное утверждение:

*Пусть даны два определения. Пусть обобщенное решение существует в смысле определения 1 и единственно в смысле определения 2. Пусть, кроме того, определение 2 имеет вид: "Под обобщенным решением будем понимать функцию, удовлетворяющую следующим свойствам...". Тогда, если обобщенное решение в смысле определения 1 удовлетворяет всем свойствам в определении 2, то определения 1 и 2 эквивалентны. При этом существует и притом единственное обобщенное решение в смысле определения 1 и существует и притом единственное обобщенное решение в смысле определения 2, которое совпадает с обобщенным решением в смысле определения 1. Кроме того, если в смысле одного из определений обобщенное решение устойчиво по начальным данным, то оно будет устойчиво и в смысле другого определения.*

Это утверждение будет использоваться нами в § 1 следующим образом. Установим существование обобщенного решения в смысле Гельфанда–Lax'а (определение 1), покажем, что оно удовлетворяет определению С.Н. Кружкова (определение 7). Затем покажем, что обобщенное по С.Н. Кружкову решение удовлетворяет определению О.А. Олейник (определение 3). И, наконец, установим единственность устойчивость по начальным данным обобщенного решения в смысле О.А. Олейник. Таким образом, мы докажем эквивалентность и корректность в классе ограниченных измеримых функций этих трех базовых, часто употребляемых в литературе для случая $\eta(u)=u$, определений.

В § 2 исследуется связь уравнения (1) с уравнением Гамильтона–Якоби

$$\frac{\partial U}{\partial t}+H\left(\frac{\partial U}{\partial x}\right)=0, \text{ где } H\equiv\varphi\left(\eta^{-1}(\cdot)\right).$$

$$\frac{\partial}{\partial x}\left(\frac{\partial U}{\partial t}+H\left(\frac{\partial U}{\partial x}\right)\right)=\frac{\partial}{\partial t}\frac{\partial U}{\partial x}+\frac{\partial}{\partial x}H\left(\frac{\partial U}{\partial x}\right)=\frac{\partial v}{\partial t}+\frac{\partial H(v)}{\partial x}=0,$$

где $v=\dfrac{\partial U}{\partial x}$. Приводятся достаточные условия, при которых $u=\eta^{-1}(v)=\eta^{-1}(U_x)$.

В начале § 3 рассматривается экономическая модель распределения предприятий по уровням эффективности с учетом выбытия мощностей,

$$h\frac{du^n}{dt}=-\Phi\left(u^n\right)\left(u^n-u^{n-1}\right)+\mu\cdot\left(u^{n+1}-u^n\right),\ u^n(0)=u_0(nh),$$

где $u^n(t)$ – доля предприятий, находящихся в момент времени $t$ на уровнях не выше чем $n$. При условиях: $u_0(x)$ – функция распределения, которая имеет ограниченные производные до второго порядка включительно;

$$\Phi(u)>0,\quad \mu\ge 0,\quad \Phi'(u)>0,\quad \eta'(u)=\frac{1}{\Phi(u)+\mu},\quad \varphi'(u)=\frac{\Phi(u)-\mu}{\Phi(u)+\mu},$$

удалось показать, что

$$\forall\, n\in N,\ t\in[0,T]\to\left|u^n(t)-u(t,nh)\right|\le TO\left(\sqrt[4]{h}\right).$$

В конце § 3 показано, что если $\mu=0$, то при гладкой функции $\Phi(u)>0$ дифференциально-разностное уравнение Полтеровича–Хенкина, описывающие динамику распределения предприятий по уровням эффективности

$$h\frac{du^n}{dt} = -\Phi\left(u^n\right)\left(u^n - u^{n-1}\right),$$

устойчиво аппроксимирует (1) в $L_1$.

Заметим, что при

$$\Phi'\left(u\right) = \left(\frac{\varphi'\left(u\right)}{\eta'\left(u\right)}\right)' = \varphi''\left(\eta^{-1}\left(y\right)\right)\Big|_{y=\eta(u)} < 0$$

уравнение (1) может иметь разрывные обобщенные решения.

В приложении описывается метод исследования производных решения $u_\varepsilon\left(t,x\right)$ задачи Коши (5), (6), основанный на обобщенном принципе максимума для уравнений параболического типа. С помощью этого метода были получены равномерные по $\varepsilon > 0$ оценки на $\frac{\partial u_\varepsilon\left(t,x\right)}{\partial x}$, $\frac{\partial^2 u_\varepsilon\left(t,x\right)}{\partial x^2}$, необходимые в § 1, § 3.

Как уже отмечалось, подавляющая часть литературы посвящена случаю $\eta(u) = u$. Из приведенной в списке литературы, посвящённой квазилинейным уравнениям, лишь в [3] – [8] и частично в [9] рассматривается случай $\eta(u) \neq u$. Поэтому, если в последующем тексте номер ссылки не содержится в указанном списке, то это означает, что нужные утверждения работы можно обобщить на случай $\eta'(u) > 0$. В некоторых случаях авторы сами указывают на возможность обобщения всех результатов, не приводя доказательства, например, важный для нас случай статьи [10]. Обычно, обобщение сводится просто к переписыванию формул с небольшой, достаточно очевидной модификацией, и, с точки зрения новых идей, не представляет интереса. Нетривиальные примеры представляют собой статьи [2], [11] – [13]. Распространение некоторых результатов статей [2], [12], [13] на случай $\eta'(u) > 0$ приведено в настоящей работе. Обобщение результатов [11] на случай $\eta'(u) > 0$ содержится в работах [7], [8].

Автор благодарен чл.-корр. РАН, проф. А.А. Шананину за ценные замечания и внимательное отношение к работе, а также

д.ф.-м.н. Е.Ю. Панову, к.ф.-м.н. Н.С. Петросян, д.ф.-м.н. О.С. Розановой за ценные замечания, касающиеся § 1; к.ф.-м.н. И.В. Рублёву за ценные замечания, касающиеся § 2 и Prof. D. Serre'у за ценные замечания, касающиеся § 3.

## § 1 ОПРЕДЕЛЕНИЕ С.Н. КРУЖКОВА

Рассмотрим квазилинейное уравнение первого порядка в слое $\pi_T = \{(t,x): 0 \le t \le T, -\infty < x < +\infty\}$

$$\frac{\partial \eta(u)}{\partial t} + \frac{\partial \varphi(u)}{\partial x} = 0 \ , \tag{1.1}$$

где[2] $\eta(u)$, $\varphi(u) \in C^2$, $\eta'(u) > 0$ .Это уравнение (1.1) возникает при моделировании транспортных потоков (модель Лайтхилла–Уизема [глава 3, 1]); в газовой динамике [глава 6, 1], [3], [глава 2, 9] [глава 9, 14], [15], [16]: уравнения Эйлера, уравнения Навье–Стокса. Как правило, в физике оно означает некий закон сохранения (заряда, количества вещества, энергии), условие непрерывности [стр. 8, 17] (изменение со временем количества какой-то величины в единице объема происходит только за счет "вытекания" или "втекания" в единицу объема этой величины), и записывается чаще в виде

$$\frac{\partial \eta(u)}{\partial t} + div_x\left(\varphi(u)\right) = 0 \ .$$

Поставим начальное условие Коши

$$u\big|_{t=0} = u_0(x), \tag{1.2}$$

где $u_0(x)$ – ограниченная ($u_0(x) \in [a,b]$) измеримая функция, заданная при всех $x \in R^1$.

Для нелинейных гиперболических уравнений гладкое решение задачи Коши существует, как правило, только в малой окрест-ности линии или поверхности, где заданы начальные условия. По разрывным начальным условиям решение задачи Коши для нелинейных уравнений, вообще говоря, не определяется однозначно даже в сколь угодно малой окрестности линии или поверхности, где заданы начальные условия. Для того чтобы задача Коши для нелинейных уравнений с гладкими или разрывными начальными условиями была однозначно разрешима в большей области, необходимо рассматривать разрывные решения уравнения и по-новому поставить задачу Коши [18], [§ 9, 19], [20].

---

[2] Если $\eta'(u) < 0$, то, делая замену $\tilde{\eta}(u) = -\eta(u)$, $\tilde{\varphi}(u) = -\varphi(u)$, снова придем к $\tilde{\eta}'(u) > 0$. Исследованием уравнения (1.1) в случае, когда $\eta(u) = u$, $\varphi(u) \in C$ занимались С.Н. Кружков, Ph. Bénilan и Е.Ю. Панов.

Основная идея заключается в том, что равенства (1.1), (1.2) надо понимать в слабом смысле, т.е. для любой гладкой финитной функции $f(t,x)$ выполняется

$$0=\iint_{\pi_\infty}\left\{\frac{\partial\eta}{\partial t}+\frac{\partial\varphi}{\partial x}\right\} f\, dt\, dx=$$

$$=\int_{R^1}\eta(u_0(x))\, f(0,x)\, dx-\iint_{\pi_\infty}\left\{\frac{\partial f}{\partial t}\eta+\frac{\partial f}{\partial x}\varphi\right\} dt\, dx. \qquad (1.3)$$

Отметим, что это интегральное равенство аналогично тем соотношениям, с помощью которых вводится обобщенное решение в смысле Соболева [§ 5, 21].

Получим условие на разрыве в предположении, что начальная функция имеет вид ступеньки, которая равна $u_-$ слева и $u_+$ справа. Для этого перепишем (1.3) используя формулу Грина. Причем область интегрирования в (1.3), ограниченную контуром $\Gamma$, выберем так, чтобы она содержала разрыв, т.е. содержала часть кривой $L$ на плоскости $t$, $x$, вдоль которой решение терпит разрыв

$$\oint_{\Gamma}\left(-\eta(u)dx+\varphi(u)dt\right)=0. \qquad (1.4)$$

Пусть, кроме того, контур $\Gamma$ взят в виде узкого прямоугольного четырехугольника, ширина которого настолько мала по сравнению с длиной, что интегралом по участкам контура $\Gamma$, поперечным к $L$, можно пренебречь (рис. 1). Тогда из (1.4) следует, что

$$0=\oint_{\Gamma}\left(\eta(u)dx-\varphi(u)dt\right)=$$

$$=\left(\eta(u_+)k-\varphi(u_+)\right)\Delta t-\left(\eta(u_-)k-\varphi(u_-)\right)\Delta t+o(\Delta t)\,, \qquad (1.5)$$

где $k=\dfrac{dx}{dt}$ – соответствует наклону касательной к $L$, $\Delta t$ – длина проекции контура на ось $t$. При $\Delta t\to 0$ равенство (1.5) переходит в следующее условие для скорости движения разрыва, которое для случая $\eta(u)=u$ называется условием Гюгонио–Ренкина [стр. 42, 17] (это условия также называют условием Гюгонио [§ 64, 22])

$$k=\frac{\varphi(u_+)-\varphi(u_-)}{\eta(u_+)-\eta(u_-)} \text{ (R–H).} \qquad (1.6)$$

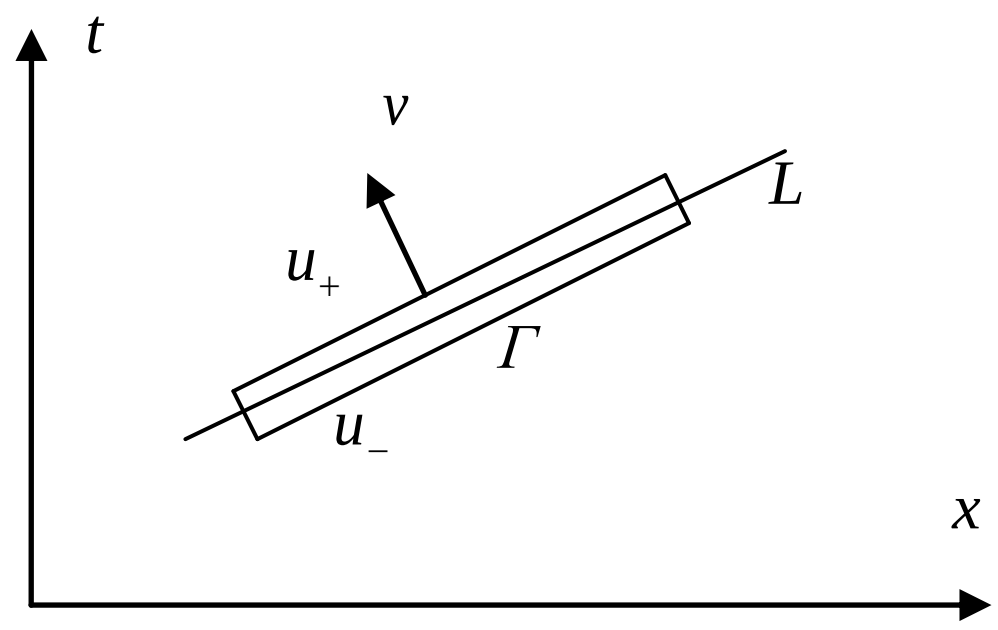

Рис. 1

Заметим, что уравнение (1.4), где $\eta\left(u\left(t,x\right)\right)$ – плотность вещества в точке $x$ в момент времени $t$, а $\varphi\left(u\left(t,x\right)\right)$ – величина потока вещества (за положительное направление потока принимается направление слева направо) в точке $x$ в момент времени $t$, для любого прямоугольного контура Г в полуплоскости $\pi_\infty$, со сторонами параллельными осям, получается очевидным образом из следующего условия сохранения массы (количества вещества)

$$\int_a^b \eta\left(u\left(t+\Delta,x\right)\right)dx-\int_a^b \eta\left(u\left(t,x\right)\right)dx=$$

$$= -\left\{ \int_t^{t+\Delta} \varphi\big(u(\tau,b)\big)d\tau - \int_t^{t+\Delta} \varphi\big(u(\tau,a)\big)d\tau \right\}.$$

Условие, что контур прямоугольный, со сторонами параллельными осям, не является существенным, поскольку любой кусочно-гладкий контур может быть аппроксимирован рассмотренными.

Оказывается, что уравнение (1.3) всегда имеет решение [6], но, как показывает следующий пример, оно может иметь бесконечно много решений, т.е. нет единственности.

**Пример (О.А. Олейник).** Рассмотрим уравнение E. Hopf'a [стр. 80, 17], [20]

$$\frac{\partial u}{\partial t} + u\frac{\partial u}{\partial x} = 0 \tag{1.7}$$

и начальное условие типа Римана

$$u_0(x) = \begin{cases} 1, & x \le 0 \\ -1, & x > 0 \end{cases}. \tag{1.8}$$

Так, например, при любом $q \ge 1$ определенная в точках полуплоскости $t \ge 0$, функция

$$u_q(t,x) = \begin{cases} 1, & x \le \dfrac{1-q}{2}t \\ -q, & \dfrac{1-q}{2}t < x \le 0 \\ q, & 0 < x \le \dfrac{q-1}{2}t \\ -1, & \dfrac{q-1}{2}t < x \end{cases}$$

удовлетворяет при $t \ge 0$ (1.1), (1.2) в смысле (1.3) (достаточно проверить, что на разрывах выполняется условие R–H).

Отметим также, что классический метод характеристик для решения уравнений в частных производных первого порядка может использоваться лишь локально для уравнения (1.1), т.к. по прошествии некоторого времени характеристики могут начать пе-ресекаться и возникнет неоднозначность: одной точке $(t, x)$ будут соответствовать несколько, вообще говоря, разных значений $u$ [стр. 24 – 28, 1] , [стр. 16 – 18, 17]. Собственно, там, где характеристики начинают пересекаться, и возникает разрыв у решения (1.1). Метод характеристик вкупе с условиями на разрыве был одним из первых методов исследования задачи Коши (1.1), (1.2) [4], [23].

Проблема корректного определения обобщенного решения квазилинейных уравнений возникла в начале 50-ых и изучалась в работах И.М. Гельфанда, О.А. Олейник, P.D. Lax'а, А.Н. Тихонова, А.А. Самарского, О.А. Ладыженской. Общая идея состоит в том, что наряду с уравнением (1.1) предлагается рассматривать "возмущенное" уравнение[3]

$$\frac{\partial \eta(u_\varepsilon)}{\partial t} + \frac{\partial \varphi(u_\varepsilon)}{\partial x} = \varepsilon \frac{\partial^2 u_\varepsilon}{\partial x^2}\ , \ \varepsilon > 0, \tag{1.9}$$

и понимать решение задачи Коши (1.1), (1.2) как предел при $\varepsilon \to 0+$ решений задач Коши (1.9), (1.2).

В [10] показано, что при произвольной ограниченной измеримой начальной функции существует решение задачи Коши (1.1), (1.2), понимаемое как предел при $\varepsilon \to 0+$ решений задач Коши (1.9), (1.2), причем равенство (1.2) понимается в слабом смысле.

Идея доказательства этого утверждения следующая [10]. С помощью ядра усреднения ("шапочек Ляпунова") [10], [12], модуля непрерывности, свойств точек Лебега [10], [стр. 236, 24] и обобщенного принципа максимума [4], [5], [7], [10] устанавливается компактность по норме $L_1$ по теореме Колмогорова–Тулайкова [10], [стр. 457, 24], [§ 1 главы 9, 25] последовательности по $\varepsilon$ $\{u_\varepsilon(t, x)\}_{\varepsilon>0}$ – решений задач Коши (1.9), (1.2). Далее, с помощью диагонального процесса выделяется подпоследовательность, которая сходится почти всюду в $\pi_T$. Затем в интегральном законе со-

---

[3] В современной литературе его называют законом сохранения с искусственной вязкостью или вязким законом сохранения [15]. Интерес к уравнениям такого типа был вызван пионерской работой E. Hopf'а [20].

хранения переходим к пределу под знаком интеграла по теореме Лебега об ограниченной сходимости. Если начальная функция ограниченна, то, по обобщенному принципу максимума, этой же константой будут равномерно по $\varepsilon > 0$ ограничены функции $\{u_\varepsilon(t,x)\}_{\varepsilon>0}$ (см. замечание 1 приложения, а также [4], [7], [8]). Поэтому, предельная функция будет удовлетворять интегральному закону сохранения (1.3). Аналогичным образом проверяем выполнение начальных условий в слабом смысле. Таким образом, доказано существование решения, понимаемого в смысле следующего определения:

**Определение 1 (И.М. Гельфанд, P.D. Lax).** *Ограниченную измеримую функцию $u(t,x)$ будем называть допустимым решением задачи Коши для уравнения (1.1) в $\pi_\infty$ с начальным условием (1.2), если $u(t,x) = \lim_{\varepsilon\to 0+} u_\varepsilon(t,x)$ - есть предел почти всюду по $x$ последовательности по $\varepsilon$ $\{u_\varepsilon(t,x)\}_{\varepsilon>0}$ при $\varepsilon \to 0+$ и любом фиксированном $t > 0$, где $u_\varepsilon(t,x)$ – решение задачи Коши (1.9), (1.2), с теми же функциями $\eta(u)$ и $\varphi(u)$, что и в уравнении (1.1), а нижний индекс $\varepsilon$ у решения $u_\varepsilon(t,x)$ соответствует коэффициенту $\varepsilon$ при второй производной по $x$ в уравнении (1.9).*

В приводимых ниже определении 2, утверждениях 1, 2 предполагается, что $\varphi\left(\eta^{-1}(y)\right)$ – строго выпуклая функция, т.е. $\left(\varphi'(u)/\eta'(u)\right)' > 0$ (случай когда $\varphi\left(\eta^{-1}(y)\right)$ – строго вогнутая функция, сводится к строго выпуклому заменой $\tilde{x} = -x$ и $\tilde{\varphi}(u) = -\varphi(u)$ ).

Изучая разностную схему предложенную P.D. Lax'ом [18] для (1.1) и (1.9), О.А. Олейник для случая $\eta(u) = u$ обнаружила следующее свойство решения задачи Коши (1.9), (1.2) [12] (для случая $\eta'(u) > 0$ см. утверждение 2 приложения, а также [7]).

**Утверждение 1.** *Пусть* $\left(\varphi'(u)/\eta'(u)\right)' > 0$. *Тогда для всех* $x$ *и* $t \ge t_0 > 0$

$$\frac{\partial u_\varepsilon}{\partial x} \le \frac{E}{t} \quad ,$$

*где $E$ не зависит от $\varepsilon$.*

**Определение 2 (О.А. Олейник).** *Ограниченную измеримую функцию $u(t,x)$ будем называть обобщенным по О.А. Олейник решением задачи Коши (1.1), (1.2) в $\pi_\infty$, если*

1) *для всякой финитной непрерывно дифференцируемой в $\pi_\infty$ функции $f(t,x)$, почти всюду выполняется равенство (1.3).*
2) $\forall t_0 > 0 : \exists E : \forall t > t_0 > 0;\ x_1, x_2 \in R^1$

$$\frac{u(t,x_1) - u(t,x_2)}{x_1 - x_2} \le \frac{E}{t},$$

*если $u(t,x)$ изменить, быть может, на множестве меры нуль.*

Заметим, что условие пункта 2 определения 2 обеспечивает единственность обобщенного решения задачи Коши в примере Олейник.

Оказывается, что и в общем случае при $\eta(u) = u$ обобщенное по О.А. Олейник решение задачи Коши (1.1), (1.2) существует и единственно [4], [12]. Для $\eta'(u) > 0$ доказательство единственности аналогично [12]. Поэтому, из утверждения 1 получаем эквивалентность определений 1 и 2, когда $(\varphi'(u)/\eta'(u))' > 0$.

Рассмотрим теперь более подробно преобразования уравнения (1.1), приведённые в введении.

1) Глядя на уравнение (1.1), естественно возникает желание разделить его на $\eta'(u) \ge a > 0$ и, таким образом, привести к виду

$$\frac{\partial u}{\partial t} + \frac{\varphi'(v)}{\eta'(u)} \frac{\partial u}{\partial x} = 0 \ . \qquad (1.10)$$

Однако, формы записи (1.1) и (1.10) будучи эквивалентны с точки зрения гладких решений, не будут таковыми для разрывных решений.

Проиллюстрируем это на следующем примере И.М. Гельфанда [2]. Возьмем уравнение E. Hopf'а (1.7)

$$\frac{\partial u}{\partial t}+u\frac{\partial u}{\partial x}=0 \ ,$$

и запишем его в двух дивергентных формах:

$$\frac{\partial u}{\partial t}+\frac{\partial (u^2/2)}{\partial x}=0,$$

$$\frac{\partial (u^2/2)}{\partial t}+\frac{\partial (u^3/3)}{\partial x}=0.$$

Написав для каждого из этих уравнений условие (1.6), легко убеждаемся в неэквивалентности.

Приводимое ниже утверждение 2 при $(\varphi'(u)/\eta'(u))'>0$ отвечает на вопрос: какие скорости движения одного и того же разрыва можно получить, умножая уравнение (1.1) на различные положительные, достаточно гладкие функции $\psi(u)$?

**Утверждение 2.** *Пусть* $(\varphi'(u)/\eta'(u))'>0$. *Тогда следующие два условия равносильны*

*1)* $\exists\,\psi(u)>0$ *– положительная гладкая функция такая, что уравнение (1.1), умноженное на* $\psi(u)$, *будет иметь скорость разрыва* $k$.

*2)* $$\frac{\varphi'(u_+)}{\eta'(u_+)}<k<\frac{\varphi'(u_-)}{\eta'(u_-)}. \qquad (1.11)$$

**Доказательство.** После умножения на $\psi(u)$ (1.1), запишем то, что получится в дивергентном виде (т.е. занесем функции, стоящие перед частными производными, под соответствующие частные производные)

$$\frac{\partial f(u)}{\partial t}+\frac{\partial g(u)}{\partial x}=0 \ .$$

Причем, в силу тех условий, которые мы наложили на $\psi(u)$, $f(u)$ – строго монотонная, возрастающая и достаточно гладкая функция (как композиция функций этого класса). Также, очевидно, что

$$\frac{g'(u)}{f'(u)} = \frac{\varphi'(u)}{\eta'(u)}. \quad (1.12)$$

Из (1.12) имеем

$$g'(u) = \frac{\varphi'(u)}{\eta'(u)} f'(u),$$

следовательно

$$g(u_+) - g(u_-) = \int_{u_-}^{u_+} g'(u)\,du = \int_{u_-}^{u_+} \frac{\varphi'(u)}{\eta'(u)} f'(u)\,du. \quad (1.13)$$

По формуле R–H находим, с учетом (1.13), что получившееся при умножении уравнение имеет скорость движения разрыва

$$k = \frac{\int_{u_-}^{u_+} \frac{\varphi'(u)}{\eta'(u)} f'(u)\,du}{f(u_+) - f(u_-)} = \frac{\int_{u_+}^{u_-} \frac{\varphi'(u)}{\eta'(u)} f'(u)\,du}{f(u_-) - f(u_+)}. \quad (1.14)$$

Из того, что $f(u)$ – строго монотонно возрастающая, достаточно гладкая функция и выполняется (1.14) следует (1.11). Утверждение 2 доказано.

2) Также возникает желание сделать замену[4] $v = \eta(u)$, которая приведет (1.1) к хорошо изученному виду

$$\frac{\partial v}{\partial t} + \frac{\partial \varphi\left(\eta^{-1}(v)\right)}{\partial x} = 0 . \quad (1.15)$$

Так как под решением (1.1) мы можем понимать предел почти всюду при $\varepsilon \to 0+$ решений (1.9) с той же начальной функцией и теми же коэффициентами при частных производных, то для (1.15) получим, что $v=\lim\limits_{\varepsilon\to 0+} v_\varepsilon$ – предел почти всюду по $x$, при произвольном фиксированном $t>0$, где $v_\varepsilon$ – решение

$$\frac{\partial v_\varepsilon}{\partial t}+\frac{\partial\varphi\left(\eta^{-1}\left(v_\varepsilon\right)\right)}{\partial x}=\varepsilon\frac{\partial^2 v_\varepsilon}{\partial x^2},$$

с начальной функцией $v_\varepsilon(0,x)=\eta(u_0(x))$.

[4] Собственно такую замену и предлагала сделать О.А. Олейник в работах [4], [23], [26]. Следует отметить, что это предлагалось сделать до того, как определялось обобщенное решение, поэтому не было необходимости обосновывать возможность проведения такой замены.

Сопоставляя (1.9), которое с помощью замены $v_\varepsilon=\eta(u_\varepsilon)$ приводится к виду (1.15) с дивергентной вязкостью [2]

$$\frac{\partial v_\varepsilon}{\partial t}+\frac{\partial\varphi\left(\eta^{-1}\left(v_\varepsilon\right)\right)}{\partial x}=\varepsilon\frac{\partial^2\eta^{-1}(v_\varepsilon)}{\partial x^2}=\varepsilon\frac{\partial}{\partial x}\left[B\left(v_\varepsilon\right)\frac{\partial v_\varepsilon}{\partial x}\right],$$

$$B\left(v_\varepsilon\right)=1/\eta'\left(\eta^{-1}\left(v_\varepsilon\right)\right)>0,$$

$$v_\varepsilon(0,x)=\eta(u_0(x)),$$

$$u=\lim_{\varepsilon\to 0+}u_\varepsilon=\lim_{\varepsilon\to 0+}\eta^{-1}(v_\varepsilon)=\eta^{-1}\left(\lim_{\varepsilon\to 0+}v_\varepsilon\right),$$

и

$$\frac{\partial v_\varepsilon}{\partial t}+\frac{\partial\varphi\left(\eta^{-1}\left(v_\varepsilon\right)\right)}{\partial x}=\varepsilon\frac{\partial^2 v_\varepsilon}{\partial x^2},$$

$$v_\varepsilon(0,x)=\eta(u_0(x)),$$

$$u=\lim_{\varepsilon\to 0+}u_\varepsilon=\lim_{\varepsilon\to 0+}\eta^{-1}(v_\varepsilon)=\eta^{-1}\left(\lim_{\varepsilon\to 0+}v_\varepsilon\right),$$

получим, что, вообще говоря, непонятно, связаны ли допустимые решения (1.1) и (1.15) соотношением $v=\eta(u)$.

**Замечание.** Приведенные выше рассуждения поясняют связь между задачами исследования поведения допустимого решения (1.1) и исследования квазилинейного уравнения параболического типа с дивергентной вязкостью при $\varepsilon\to 0+$ [2]

$$\frac{\partial v_\varepsilon}{\partial t}+\frac{\partial\varphi\left(\eta^{-1}\left(v_\varepsilon\right)\right)}{\partial x}=\varepsilon\frac{\partial}{\partial x}\left[B\left(v_\varepsilon\right)\frac{\partial v_\varepsilon}{\partial x}\right],$$

$B\left(v_\varepsilon\right)>0$ (задача И.М. Гельфанда).

Гипотеза, высказанная И.М. Гельфанда в работе [2], в частности, говорит, что $\lim_{\varepsilon\to 0+} v_\varepsilon$ не зависит от $B(v_\varepsilon)>0$ и, как следствие, допустимые решения (1.1) и (1.15) связаны соотношением $v=\eta(u)$. Ниже (теоремы 1, 2) будет показана справедливость гипотезы И.М. Гельфанда. Отметим также связь задачи И.М. Гельфанда с изучением одного из уравнений системы Навье–Стокса [15], [16].

На практике оказалось удобным рассматривать кусочно-непрерывные и кусочно-гладкие начальные условия с конечным числом точек разрыва, т.к. обобщенные решения непрерывно зависят от начальных данных, а любую ограниченную измеримую функцию можно приблизить функциями, указанного выше класса. Оказывается, что тогда функция $u(t,x)$ – непрерывно дифференцируемая всюду в полуплоскости $\pi_\infty$, кроме конечного числа кусочно-гладких линий, в точках которых может иметь разрывы, при этом во всех точках разрыва, за исключением, быть может, конечного числа, существуют пределы справа и слева. Это побудило О.А. Олейник, ограничившись указанным выше классом начальных функций, дать в конце 50-ых следующее конструктивное определение обобщенного решение задачи Коши (1.1), (1.2) [27].

**Определение 3 (О.А. Олейник).** *Ограниченную измеримую функцию* $u(t,x)$ *будем называть обобщенным решением задачи Коши (1.1), (1.2) в* $\pi_\infty$, *если*

1) *удовлетворяет (1.3) почти всюду;*
2) *на разрыве удовлетворяет R–H условию;*
3) *на разрыве выполняется E – условие*

   $\forall\, u\in(u_-,u_+)\to\sigma(u_-,u_+)\le\sigma(u_-,u)$, *если* $u_-<u_+$,

   $\forall\, u\in(u_+,u_-)\to\sigma(u_-,u_+)\ge\sigma(u_-,u)$, *если* $u_->u_+$,

   *которое впервые было введено О.А. Олейник в [28], как условие устойчивости (допустимости) разрыва.*

Это определение оказалось очень полезным. Например, оно широко использовалось в 70-ые, 80-ые годы школой T.P. Liu [29] – [31]. В работах [29] – [31] исследовалось асимптотическое по вре-мени поведение решения задачи Коши (1.1), (1.2) в случае не выпуклой и невогнутой функции $\varphi\left(\eta^{-1}(y)\right)$, при этом исследование частично сводится к решению задачи о распаде разрыва [2], [стр. 85 – 90, 17], точнее к решению задачи об одновременном распаде конечного числа разрывов [31]. Конструктивное определение 3 позволяет явно решать такого рода задачи.

Как было показано в [27], условия 2), 3) определения 3 обеспечивают единственность и устойчивость в слабом смысле по начальным данным обобщенного решения, когда $\eta(u)=u$. Для $\eta'(u)>0$ доказательство единственности и устойчивости аналогично [27]. В [2], [стр. 62 – 67, 17] для случая $\eta(u)=u$ и в [стр. 581 – 582, 9] для случая $\eta'(u)>0$ показывается, что для начальных условий типа Римана (существуют конечные пределы $\lim\limits_{x\to\pm\infty} u_0(x)$) допустимое решение Гельфанда–Lax'а удовлетворяет условиям 2) и 3) определения 3. В конце 50-ых установлением связи между определениями 1 и 3, при $\eta(u)=u$ и вязкостном слагаемом в (1.9) вида $t\varepsilon\dfrac{\partial^2 u}{\partial x^2}$, занимался А.С. Калашников [32].

Отметим еще одно определение (о слабой замкнутости или об устойчивости в "потенциальной метрике"), данное О.А. Олейник [27], [28] в конце 50-ых, и основанный на этом определении численный метод Б.Л. Рождественского [33], [стр. 564 – 566, 9] решения задачи (1.1), (1.2) – "метод потенциального сглаживания". Связь определений 3 и 4 изучалась для случая $\eta(u)=u$ в [27], [28], [стр. 510 – 522, 9]. Далее везде будем считать, что определение 4 даётся при условии $\eta(u)=u$.

**Определение 4 (О.А. Олейник, Б.Л. Рождественский).** *Ограниченная измеримая функция $u(t,x)$, удовлетворяющая почти всюду интегральному закону сохранения (1.3), называется слабо устойчивым (говорят также потенциально устойчивым [33]) решением задачи Коши (1.1), (1.2) в слое $\pi_T$, если*

$$\text{из} \sup_{-\infty\le a<b\le\infty} \left| \int_a^b \left(u_0^k(x) - u_0(x)\right) dx \right| \xrightarrow{k\to\infty} 0 \Rightarrow$$

$$\forall\, 0 \le t \le T \sup_{-\infty\le a<b\le\infty} \left| \int_a^b \left(u^k(t,x) - u(t,x)\right) dx \right| \xrightarrow{k\to\infty} 0,$$

*где* $u^k(t,x)$ *– решение (1.1) с начальным условием Коши* $u^k(t,0) = u_0^k(x)$.

Недавно Е.Ю. Пановым был предложен контрпример, который показывает, что определение 4 при $\eta'(u) > 0$ не эквивалентно основному определению 7. А при $\eta(u) = u$ доказана теорема[5] о том, что слабый предел последовательности энтропийных решений будет энтропийным решением.

Из эквивалентности определений 3 и 4 (см. теорему 1) следует, что условие (1.11), где $k$ находится из условия R–H, обеспечивают в случае строго выпуклой функции $\varphi(u)$ ($\eta(u) = u$) устойчивость решения (1.1), (1.2) по начальным данным. Действительно, если $\varphi(u)$ – строго выпукла, то Е – условие переписывается в виде (1.11), где $k$ находится из условия R–H. Кроме того, в случае строго выпуклой функции $\varphi(u)$ условие (1.11) является критерием устойчивости разрывного решения относительно возмущений самого решения [стр. 42 – 43, 1], [2], [§ 50, 22]. Условие (1.11) иногда называют условием возрастания энтропии или условием P.D. Lax'а [2], [34].

Приведем пример [стр. 510 – 513, 9], где условие устойчивости (1.11) "отбрасывает" не физическое ("надуманное") решение (1.3).

**Пример (Б.Л. Рождественский, Н.Н. Яненко).** Снова рассмотрим уравнение E. Hopf'а (1.7)

---

[5] Обобщающая результаты L. Tartar'a (1979), о том, что слабый предел последовательности энтропийных решений (см. определение 7) будет слабым решением.

$$\frac{\partial u}{\partial t} + u\frac{\partial u}{\partial x} = 0$$

и начальное условие отличное от (1.8)

$$u_0(x) = \begin{cases} -1, & x \le 0 \\ 1, & x > 0 \end{cases}.$$

Возможны следующие решения (1.3):

$$u_1(t,x) = \begin{cases} -1, & x \le -t \\ \dfrac{x}{t}, & -t \le x \le t \\ 1, & x \ge t \end{cases} \qquad u_2(t,x) = \begin{cases} -1, & x < 0 \\ 1, & x > 0 \end{cases}.$$

Размазывая разрыв начальных данных в точке $x = 0$, т.е. вводя $u_0^\delta(x)$ – монотонно возрастающую непрерывную функцию, совпадающую вне отрезка $|x| \le \delta$ с $u_0(x)$, мы увидим, что $\lim\limits_{\delta\to 0} u^\delta(t,x) = u_1(t,x)$. Т.е. неклассическое решение $u_2(t,x)$ не является устойчивым решением в смысле определения 4. Несложно проверить, что условие (1.11) не выполняется на разрыве решения $u_2(t,x)$: $1 \le 0 \le -1$, т.е. не выполняется Е – условие, поэтому $u_2(t,x)$ не является решением и в смысле определения 3. Если посмотреть на поведение характеристик системы, то можно заметить, что для решения $u_2(t,x)$ разрыв надуман – он не вызван пересечением характеристик. Вместо того чтобы пересекаться на разрыве, характеристики расходятся от разрыва [стр. 512, 9].

Заметим, что процесс, описываемый разрывным решением (1.1), не обратим во времени [стр. 67 – 68, 17], [стр. 519 – 522, 9]. Причем условие разрывности процесса существенно для необратимости. Так в примере О.А. Олейник, приведенном ближе к началу этого параграфа, $u_q(t,x)$ при $q = 1$ является разрывным решением (1.7), для которого выполняется Е – условие, для строго выпуклой функции $\varphi\left(\eta^{-1}(u)\right) = \varphi(u) = u^2/2$, т.е. $u_q(t,x)$ при $q = 1$ являет-ся устойчивым решением в смысле определения 3. Если решать (1.7) в попятном времени, то неравенства в Е – условии поменяются на противоположные, поэтому, если $u_q(t,x)$ при $q=1$ удовлетворяла прямому Е – условию, то оно точно не может удовлетворять попятному Е – условию.

Уже в 50-ые годы было известно, что единственность решения уравнения (1.3) обеспечивают также дополнительные условия в виде интегральных неравенств, которые в гидромеханике и газовой динамике интерпретируются как условия возрастания энтропии, условие диссипации энергии [3], [стр. 67 – 72, 17], [§ 54, 22], [§ 17, 35], [§ 29, 36], [37], [стр. 75, 38]. В работе [39] С.Н. Кружков получил, что условие устойчивости в $L_1$ по начальным данным:

$$\int_{a-ct}^{b+ct} \left|u_0^k(x)-u_0(x)\right| dx \xrightarrow{k\to\infty} 0 \;\Rightarrow\; \int_a^b \left|u^k(t,x)-u(t,x)\right| dx \xrightarrow{k\to\infty} 0,$$

$$c=\max_{u\in[-M,M]}\left|\varphi'(u)/\eta'(u)\right|$$

может быть записано в виде интегрального (подобного (1.3)) неравенства, точнее в виде параметрического семейства интегральных неравенств, которые включают в себя условия возрастания энтропии. Впоследствии, в работе [10] С.Н. Кружков подробно исследовал данное им определение.

Далее для нас будет удобно представить (1.9) в виде

$$0=\varepsilon\frac{\partial^2 u}{\partial x^2}-\frac{\partial\eta(u)}{\partial t}+\frac{\partial\varphi(u)}{\partial x}. \tag{1.16}$$

Пусть $\Phi(u)$ – произвольная дважды гладкая выпуклая функция, $f(t,x)\geq 0$ – произвольная дважды гладкая финитная в слое $\pi_T$ пробная функция (носитель $f(t,x)$ расположен строго внутри слоя $\pi_T$). Умножим обе части (1.16) на $\Phi'(u)f(t,x)$ и проинтегрируем по $\pi_T$, перебрасывая производные на пробную функцию. Покажем это поподробнее на примере $\dfrac{\partial\eta(u)}{\partial t}$:

$$\forall\, k \to \iint\limits_{\pi_T} \Phi'(u)\eta'(u)\frac{\partial u}{\partial t}\, f\, dx\, dt = \int\limits_{R_x^1}\left(\int\limits_k^u \Phi'(\tilde u)\eta'(\tilde u)d\tilde u\right) f\, dx\Bigg|_{t=0}^{t=T} -$$

$$-\iint\limits_{\pi_T}\left(\int\limits_k^u \Phi'(\tilde u)\eta'(\tilde u)d\tilde u\right) f_t\, dx\, dt = -\iint\limits_{\pi_T}\left(\int\limits_k^u \Phi'(\tilde u)\eta'(\tilde u)d\tilde u\right) f_t\, dx\, dt,$$

в силу финитности $f(t,x)$ в $\pi_T$.

В итоге получим

$$0 = \iint\limits_{\pi_T}\Bigg\{\int\limits_k^u \Phi'(u)\eta'(u)\,du\; f_t + \int\limits_k^u \Phi'(u)\varphi'(u)\,du\; f_x +$$

$$+\varepsilon\left(\frac{\partial^2\Phi(u)}{\partial x^2} - \Phi''(u){u_x}^2\right) f\Bigg\}dxdt,$$

$$\varepsilon\iint\limits_{\pi_T}\Phi''(u){u_x}^2 f\, dx\, dt =$$

$$= \iint\limits_{\pi_T}\left\{\int\limits_k^u \Phi'(u)\eta'(u)\,du\; f_t + \int\limits_k^u \Phi'(u)\varphi'(u)\,du\; f_x + \varepsilon\frac{\partial^2\Phi(u)}{\partial x^2} f\right\}dxdt,$$

$$0 \le \iint\limits_{\pi_T}\left\{\int\limits_k^u \Phi'(u)\eta'(u)\,du\; f_t + \int\limits_k^u \Phi'(u)\varphi'(u)\,du\; f_x + \varepsilon\frac{\partial^2\Phi(u)}{\partial x^2} f\right\}dxdt,$$

$$0 \le \iint\limits_{\pi_T}\left\{\int\limits_k^u \Phi'(u)\eta'(u)\,du\; f_t + \int\limits_k^u \Phi'(u)\varphi'(u)\,du\; f_x + \varepsilon\Phi(u) f_{xx}\right\}dxdt. \quad (1.17)$$

Устремив $\varepsilon \to 0+$, мы получим

$$0 \le \iint\limits_{\pi_T}\left\{\int\limits_k^u \Phi'(u)\eta'(u)\,du\; f_t + \int\limits_k^u \Phi'(u)\varphi'(u)\,du\; f_x\right\}dxdt, \qquad (1.18)$$

которое справедливо для любого $k$. Заметим, что в случае, когда $\Phi(u)$ линейная функция в (1.17) и (1.18) вместо неравенств можно писать равенства.

Введём обозначения:

$$\tilde{\eta}(u) = \int_k^u \Phi'(u)\eta'(u)\,du \text{ и } q(u) = \int_k^u \Phi'(u)\varphi'(u)\,du.$$

**Определение 5 (E. Jouguet, С.Н. Кружков, P.D. Lax).** *Ограниченная измеримая функция* $u(t,x)$ *называется решением задачи (1.1), (1.2) с допустимой энтропией* $\Phi(u)$ ($\Phi$ *– решением), если:*

1) $u(t,x)$ *удовлетворяет (1.3) почти всюду*
2) *(энтропийное неравенство [3]) В* $\pi_\infty$

$$\frac{\partial \tilde{\eta}(u)}{\partial t} + \frac{\partial q(u)}{\partial x} \le 0$$

*в слабом смысле, т.е. для любой гладкой финитной в* $\pi_\infty$ *функции* $f(t,x) \ge 0$ *выполняется (1.18).*

Энтропийные неравенства для уравнений газовой динамики, впервые рассматривал E. Jouguet (конец XIX начало XX вв. см. [3]). В конце 60-ых С.Н. Кружков с помощью энтропийных неравенств построил вполне законченную теорию обобщенных решений квазилинейного уравнения типа закона сохранения. В начале 70-ых исследованием энтропийных неравенств занимался P.D. Lax. В последние годы в С.К. Годунов изучает близкие вопросы. В конце 80-ых С.Н. Кружковым был поставлен вопрос: когда $\Phi$ – решение единственно? Ответ на этот вопрос был получен его учеником Е.Ю. Пановым в работе [40]. Оказывается, что в случае, когда $\varphi\left(\eta^{-1}(u)\right)$ – строго выпуклая или вогнутая функция, $\Phi$ – решение единственно. Если $\varphi\left(\eta^{-1}(u)\right)$ не является строго выпуклой или вогнутой функцией, то $\Phi$ – решение задачи Коши (1.1), (1.2) при подходящем выборе начальных данных не единственно [40]. Далее везде будем считать, что определение 5, также как и определение 2, даётся при условии $\left(\varphi'(u)/\eta'(u)\right)' > 0$.

**Определение 6 (С.Н. Кружков).** *Ограниченная измеримая функция* $u(t,x)$ *называется энтропийным решением задачи (1.1),*

*(1.2), если $u(t,x)$ – есть $\Phi$ – решение для любой дважды гладкой, выпуклой функции $\Phi(u)$.*

Это определение играет важную роль в теории квазилинейных уравнений. Приведём его к более удобному виду. Для этого покажем, что (1.18) можно записать в эквивалентной форме

$$\iint_{\pi_T} \{ sign(u(t,x)-k)\left[\eta(u(t,x))-\eta(k)\right] f_t +$$

$$+sign(u(t,x)-k)\left[\varphi(u(t,x))-\varphi(k)\right] f_x \} \, dx\, dt \geq 0. \qquad (1.19)$$

Действительно, функцию $|u-k|$ можно аппроксимировать дважды гладкими выпуклыми функциями $\Phi(u)$, поэтому по теореме Лебега об ограниченной сходимости из (1.18) следует (1.19). Пусть для любого $k$ справедливо (1.19), покажем, что отсюда следует (1.18). Для этого нам потребуется вспомогательное утверждение 3 [40].

**Утверждение 3.** *Для любой дважды гладкой функции $\Phi(u)$ справедливы представления*[6]

$$\Phi(u) = \frac{1}{2}\int_a^b |u-k|\Phi''(k)\,dk + \frac{1}{2}\left(\Phi'(a)+\Phi'(b)\right)u +$$

$$+\frac{1}{2}\left(\Phi(a)+\Phi(b)-a\Phi'(a)-b\Phi'(b)\right),$$

$$\Phi'(u) = \frac{1}{2}\int_a^b sign(u-k)\Phi''(k)\,dk + \frac{1}{2}\left(\Phi'(a)+\Phi'(b)\right),$$

*где $u \in (a,b)$.*

**Доказательство.** Приведем конструктивное доказательство этого утверждения. Необходимо определить неизвестные функции $g(k)$ и $C(u)$ ($C(u)$ – линейная функция) в разложении

[6] Для выпуклой функции $\Phi(u)$ это можно проинтерпретировать так: с точностью до линейной функции любая выпуклая дважды гладкая функция раскладывается по системе функций $|u-k|$, где $k$ пробегает некоторый отрезок $[a,b]$, причем с неотрицательными коэффициентами.

$$\Phi(u)=\int_a^b |u-k|\,g(k)\,dk+C(u).$$

$$\Phi(u)=\int_a^b |u-k|\,g(k)\,dk+C(u)=\int_a^u (u-k)g(k)\,dk+\int_u^b (k-u)g(k)\,dk+$$

$$+C(u)=u\left\{\int_a^u g(k)\,dk-\int_u^b g(k)\,dk\right\}+\int_u^b kg(k)\,dk-\int_a^u kg(k)\,dk+C(u),$$

$$\Phi'(u)=\left\{\int_a^u g(k)\,dk-\int_u^b g(k)\,dk\right\}+C'(u),$$

$$\Phi''(u)=2g(u)+C''(u)=2g(u).$$

Таким образом, мы нашли, что

$$g(u)=\frac{1}{2}\Phi''(u).$$

Но тогда,

$$\Phi(u)=\frac{1}{2}\int_a^b |u-k|\,\Phi''(k)\,dk+C(u)=\frac{1}{2}\int_a^u (u-k)\Phi''(k)\,dk+$$

$$+\frac{1}{2}\int_u^b (u-k)\Phi''(k)\,dk+C(u)=-\frac{1}{2}(u-a)\Phi'(a)+\frac{1}{2}(b-u)\Phi'(b)+$$

$$+\frac{1}{2}\int_a^u \Phi'(k)\,dk-\frac{1}{2}\int_u^b \Phi'(k)\,dk+C(u)=\frac{1}{2}\big(a\Phi'(a)+b\Phi'(b)-\Phi(a)-\Phi(b)\big)-$$

$$-\frac{1}{2}u\big(\Phi'(a)+\Phi'(b)\big)+\Phi(u)+C(u).$$

Откуда находим

$$C(u)=\frac{1}{2}\big(\Phi'(a)+\Phi'(b)\big)u+\frac{1}{2}\big(\Phi(a)+\Phi(b)-a\Phi'(a)-b\Phi'(b)\big).$$

Утверждение 3 доказано.

Умножим неравенство (1.19) на неотрицательное число $\Phi''(k)$. Посмотрим на то, что получится в правой части неравенства как на функцию от $k$. Проинтегрируем эту функцию по $k$, на некотором конечном отрезке $[a,b]$. По теореме Фубини, так как интегрирование ведется по компакту, в силу финитности $f(t,x)$, и ограниченности подынтегрального выражения на этом компакте, интеграл по $k$ существует и можно поменять местами интегрирование по $t$, $x$ и $k$.

С учетом того, что $\eta'(u) > 0$ (1.19) можно переписать как

$$\iint_{\pi_T} \left\{ \left|\eta(u(t,x)) - \eta(k)\right| f_t + sign(u(t,x)-k)\left[\varphi(u(t,x)) - \varphi(k)\right] f_x \right\} dxdt \ge 0 .$$

Таким образом, мы пришли к следующему определению, эквивалентному определению 6.

**Определение 7 (С.Н. Кружков).** *Ограниченная измеримая функция* $u(t,x)$ *называется обобщенным по С.Н. Кружкову (энтропийным [стр. 74, 17]) решением задачи (1.1), (1.2) в слое* $\pi_T$, *если:*

1) *Для любой константы* $k$ *и любой гладкой финитной в слое* $\pi_T$ *функции* $f(t,x) \ge 0$

$$\iint_{\pi_T} \left\{ \left|\eta(u(t,x)) - \eta(k)\right| f_t + \right.$$
$$\left. + sign(u(t,x)-k)\left[\varphi(u(t,x)) - \varphi(k)\right] f_x \right\} dx\,dt \ge 0 . \qquad (1.20)$$

2) $\exists\, \Theta \in [0,T] : mes(\Theta) = 0$,

*при* $t \in [0,T] \setminus \Theta$ $u(t,x)$ *определена п.в. в* $R^1$ *и*

$$\forall\, K_r = \{|x| \le r\} \subset R^1 \to \lim_{\substack{t\to 0+ \\ t\in[0,T]\setminus\Theta}} \int_{K_r} |u(t,x) - u_0(x)| = 0 .$$

Единственность и устойчивость в $L_1$ по начальному условию, так определенного, обобщенного решения, следует из [10]. Заметим, что единственность легко установить из устойчивости, “рассуждая от противного”.

Из приведенного выше построения обобщенного по С.Н. Кружкову решения ясно, что если $u(t,x)$ – является допустимым решением задачи Коши (1.1), (1.2), то $u(t,x)$ – является и обоб-щенным по С.Н. Кружкову решением задачи Коши (1.1), (1.2). Таким образом, мы получаем, что определения 1 и 7 эквивалентны.

Покажем эквивалентность определений 3 и 7. Для этого посмотрим, какому условию на разрыве должно удовлетворять обобщенное по С.Н. Кружкову решение. Пусть $\nu$ – вектор нормали к линии разрыва в точке $(t,\ x)$ (рис. 1), тогда подобно тому, как было получено условие R–H, можно получить [3], [10], что если существуют односторонние пределы слева $u_-$ и справа $u_+$, то для любого $k$ должно выполняться

$$|\eta(u_+)-\eta(k)|\cos(\nu,t)+sign(u_+-k)[\varphi(u_+)-\varphi(k)]\cos(\nu,x)\le$$

$$\le|\eta(u_-)-\eta(k)|\cos(\nu,t)+sign(u_--k)[\varphi(u_-)-\varphi(k)]\cos(\nu,x),$$

где $\cos(\nu,t)$, $\cos(\nu,x)$ – направляющие косинусы, т.е. $\cos(\nu,t)dt+\cos(\nu,x)dx=0$. Поэтому скорость движения разрыва

$$V=\frac{dx}{dt}=-\frac{\cos(\nu,t)}{\cos(\nu,x)}. \quad (1.21)$$

Для определенности будем считать, что $u_+<u_-$ (если $u_+>u_-$, то формула (1.22) будет иметь такой же вид, т.е. выполняется R–H условие, а в формулах (1.23), (1.24) надо поменять знак неравенства на противоположный). Рассмотрим два случая:

1) $k\notin[u_+,u_-]$;

2) $k\in(u_+,u_-)$.

Случай 1) дает

$$\eta(u_+)\cos(\nu,t)+\varphi(u_+)\cos(\nu,x)=\eta(u_-)\cos(\nu,t)+\varphi(u_-)\cos(\nu,x). \quad (1.22)$$

Из (1.21) и (1.22) следует, что

$$V=\frac{\varphi(u_+)-\varphi(u_-)}{\eta(u_+)-\eta(u_-)} \text{ (R–H).}$$

Случай 2) дает

$$(\eta(u_+)+\eta(u_-)-2\eta(k))\cos(\nu,t)+[\varphi(u_+)+\varphi(u_-)-2\varphi(k)]\cos(\nu,x)\le 0,$$

откуда с учетом (1.22) имеем

$$(\eta(u_+)-\eta(k))\cos(\nu,t)+[\varphi(u_+)-\varphi(k)]\cos(\nu,x)\le 0. \quad (1.23)$$

Из (1.21) и (1.23) получим, что

$$\forall\, k \in (u_+, u_-) \rightarrow \frac{\varphi(u_+) - \varphi(u_-)}{\eta(u_+) - \eta(u_-)} = V = -\frac{\cos(\nu, t)}{\cos(\nu, x)} \geq \frac{\varphi(u_+) - \varphi(k)}{\eta(u_+) - \eta(k)},$$

т.е. на разрыве обобщенного по С.Н. Кружкову решения выполняется Е – условие. Кроме того, обобщенное по С.Н. Кружкову решение удовлетворяет интегральному закону сохранения (1.3). Это сразу следует из (1.20) при $k = \pm \sup\limits_{\pi_\infty} |u(t,x)|$. Так как существует допустимое решение и любое допустимое решение удовлетворяет определению 7, то существует обобщенное по С.Н. Кружкову решение. Из того, что обобщенное по С.Н. Кружкову решение существует и удовлетворяет всем свойствам определения 3, а обобщенное решение в смысле определения 3 единственно следует, что определения 3 и 7 эквивалентны.

Подобно определениям С.Н. Кружкова, следующее определение также возникло из физических соображений.

**Определение 8 (B. Perthame, E. Tadmor).** *Функция* $u(t,x) = \lim\limits_{\varepsilon \to 0+} \int\limits_{-\infty}^{\infty} f_\varepsilon(t,x,v)\,dv$ *называется кинетическим решением задачи (1.1), (1.2), если:*

$$\exists\, f_\varepsilon(t,x,v):\ \eta'(v)\frac{\partial f_\varepsilon(t,x,v)}{\partial t} + \varphi'(v)\frac{\partial f_\varepsilon(t,x,v)}{\partial x} =$$

$$= \frac{1}{\varepsilon}\left[\chi_{u_\varepsilon(t,x)}(v) - f_\varepsilon(t,x,v)\right],\ \int\limits_{-\infty}^{\infty} f_\varepsilon(0,x,v)\,dv = u_0(x),\ \textit{где} \qquad (1.24)$$

$$\chi_u(v) = \begin{cases} sign\, u, & (u-v)v \geq 0 \\ 0, & (u-v)v < 0 \end{cases}.$$

К определению 8 можно придти, если описывать распределение плотности частиц одномерного газа $u(t,x)$ более точно (на микроскопическом уровне): а именно, считать, что плотность частиц газа $f_\varepsilon(t,x,v)$, находящихся в момент времени $t$ в точке с координатой $x$ зависит также от их скорости $v$. Заметим, что реше-ние $f_\varepsilon(t,x,v)$ кинетического уравнения типа Больцмана (1.24) представляется интегралом Дюамеля. С помощью этого представления устанавливается существование, единственность и непрерывная зависимость от начальных данных в $L_1\left(R_x^1, R_v^1\right)$ решения начальной задачи Коши для уравнения (1.24). Определение 8 для случая $\eta(u)=u$ исследовалось в начале 90-ых в работе [41]. Обобщая результаты [41], можно показать эквивалентность определений 7 и 8.

Перейдем теперь к ответу на вопрос: связаны ли обобщенные по С.Н. Кружкову решения задач Коши (1.1), (1.2) и (1.25), (1.26) соотношением $v=\eta(u)$? Сделаем в (1.1) замену $v=\eta(u)$, т.е. рассмотрим уравнение

$$\frac{\partial v}{\partial t}+\frac{\partial\varphi\left(\eta^{-1}(v)\right)}{\partial x}=0\ , \tag{1.25}$$

$$v\big|_{t=0}=\eta\left(u_0(x)\right). \tag{1.26}$$

Согласно определению 7 под обобщенным по С.Н. Кружкову решением этого уравнения будем понимать функцию, которая, помимо условия 2) определения 7, удовлетворяет следующему интегральному неравенству

$$\iint_{\pi_T}\Big\{\left|v(t,x)-k\right| f_t+$$
$$+sign\left(v(t,x)-k\right)\left[\varphi\left(\eta^{-1}\left(v(t,x)\right)\right)-\varphi\left(\eta^{-1}(k)\right)\right]f_x\Big\}dxdt\geq 0\,. \tag{1.27}$$

С другой стороны, пусть $u(t,x)$ - обобщенное по С.Н. Кружкову решение (1.1), (1.2), т.е. $u(t,x)$, помимо условия 2) определения 5, удовлетворяет (1.20). Положим в (1.20) $v=\eta(u)$

$$\iint_{\pi_T}\Big\{\left|v(t,x)-\tilde{k}\right| f_t+$$
$$+sign\left(v(t,x)-\tilde{k}\right)\left[\varphi\left(\eta^{-1}\left(v(t,x)\right)\right)-\varphi\left(\eta^{-1}(\tilde{k})\right)\right]f_x\Big\}dxdt\geq 0, \tag{1.28}$$

где $\widetilde{k}=\eta(k)$. Но так как $k$ – произвольное и $\eta'(u)>0$, то (1.28) эквивалентно (1.27).

**Утверждение 4.** *Функция* $v=\eta(u)$ *является обобщенным по С.Н. Кружкову решением*

$$\frac{\partial v}{\partial t}+\frac{\partial \varphi\left(\eta^{-1}(v)\right)}{\partial x}=0\ ,\ v(0,x)=\eta\left(u_0(x)\right),$$

*тогда и только тогда, когда* $u$ *является обобщенным по С.Н. Кружкову решением*

$$\frac{\partial \eta(u)}{\partial t}+\frac{\partial \varphi(u)}{\partial x}=0\ ,\ u(0,x)=u_0(x).$$

Сформулируем результаты, полученные в этом параграфе в виде двух теорем.

**Теорема 1.** *Определениям 1 – 8 удовлетворяет одна и та же единственная (с точностью до почти всюду по* $x$ *при фиксированном* $t>0$*) функция* $u(t,x)$*, которую будем называть обобщенным решением. Кроме того, имеет место непрерывная зависимость в* $L_1$ $u(t,x)$ *от* $u_0(x)$*:*

$$\text{если } u_0^k(x)\xrightarrow{L_1}u_0(x),\ \text{то } u^k(t,x)\xrightarrow{L_1}u(t,x).$$

Из утверждения 4 и теоремы 1, следует основной результат этой главы.

**Теорема 2.** *Функция* $v=\eta(u)$ *является обобщенным решением задачи Коши* $\dfrac{\partial v}{\partial t}+\dfrac{\partial \varphi\left(\eta^{-1}(v)\right)}{\partial x}=0$, $v(0,x)=\eta\left(u_0(x)\right)$ *тогда и только тогда, когда* $u$ *является обобщенным решением задачи Коши* $\dfrac{\partial \eta(u)}{\partial t}+\dfrac{\partial \varphi(u)}{\partial x}=0$, $u(0,x)=u_0(x)$.

Определения 1, 3, 7 – являются “базисными”, т.е. это наиболее употребляемые определения, с помощью которых порождаются практически все остальные. Почти все результаты, полученные на данный момент для квазилинейного уравнения типа закона со-хранения, базируются на одном из этих “базисных” определений. С помощью теоремы 2 можно обобщить результаты многих статей. К примеру, в гипотезе Гельфанда–Weinberg’a об асимптотическом по времени поведении обобщенного решения [2], [8], [42], [43] можно ограничиться случаем $\eta(u) = u$. И случаи, при которых гипотеза уже доказана [2], [11], [29] – [31], [44] – [50], можно обобщить на случай $\eta'(u) > 0$.[7]

## § 2 СВЯЗЬ КВАЗИЛИНЕЙНОГО УРАВНЕНИЯ ТИПА ЗАКОНА СОХРАНЕНИЯ С УРАВНЕНИЕМ ГАМИЛЬТОНА–ЯКОБИ

Рассмотрим следующую начальную задачу для уравнения Гамильтона–Якоби (далее Г–Я) с гамильтонианом $H \equiv \varphi\left(\eta^{-1}(\,\cdot\,)\right)$

$$\frac{\partial U}{\partial t} + H\left(\frac{\partial U}{\partial x}\right) = 0 \qquad (2.1)$$

$$U(0,x) = U_0(x) \;. \qquad (2.2)$$

Отметим связь уравнения (2.1) с уравнением (1.1), если $U(t,x)$ – достаточно гладкая функция:

$$\frac{\partial}{\partial x}\frac{\partial U}{\partial t} + \frac{\partial}{\partial x}H\left(\frac{\partial U}{\partial x}\right) = \frac{\partial}{\partial t}\frac{\partial U}{\partial x} + \frac{\partial}{\partial x}\varphi\left(\eta^{-1}\left(\frac{\partial U}{\partial x}\right)\right) = \frac{\partial v}{\partial t} + \frac{\partial \varphi\left(\eta^{-1}(v)\right)}{\partial x} = 0$$

$$\Leftrightarrow \; \frac{\partial \eta(u)}{\partial t} + \frac{\partial \varphi(u)}{\partial x} = 0, \text{ где } u = \eta^{-1}(U_x).$$

---

[7] Недавно были получены новые результаты, связанные с этой гипотезой:

1. Henkin G.M. Asymptotic structure for solutions of the Cauchy problem for Burgers type equations// J. fixed point theory appl., V. 1, № 2, (2007).
2. Гасников А.В. Асимптотическое по времени поведение решения начальной задачи Коши для закона сохранения с нелинейной дивергентной вязкостью// Известия РАН. Серия математическая, Т. 73, № 6, (2009).
3. Хенкин Г.М., Шананин А.А. Проблема Коши–Гельфанда и обратная задача для квазилинейного уравнения первого порядка// Функц. анализ и его прил., Т. 50, № 2, (2016).

Следует заметить, что в общем случае необходимо обосновывать возможность почленного дифференцирования уравнения (2.1). Ведь, производные решения уравнения (2.1) $U_x(t,x)$ могут терпеть разрыв первого рода. Заметим также, что разрыву производной $U_x(t,x)$ решения уравнения (2.1) будет соответствовать разрыв решения $u(t,x)$ уравнения (1.1).

Задача (2.1), (2.2), как и задача (1.1), (1.2) может не иметь классического решения. Классический метод характеристик для уравнений в частных производных первого порядка [§ 63, 51] позволяет строить решение лишь локально, вблизи поверхности, где заданы начальные условия. Сложность заключается в том, что по прошествии не большого времени характеристики начинают пересекаться. Классическим примером может служить “ласточкин хвост” $H(s)=-\sqrt{1+s^2}$, $U_0(x)=x^2/2$ [стр. 17 – 19, 52]. Уже на таком простом примере видно, что, вообще говоря, нельзя всюду требовать гладкости решения, т.е. на некотором множестве уравнение (2.1) нельзя понимать в классическом смысле. Также как и в § 1, (2.1) можно понимать в некотором “слабом” смысле. Но (2.1) вполне нелинейно и не дивергентно. Следовательно, при перебрасывании производных в уравнении (2.1) на финитную функцию $V(t,x)$, на которую предварительно умножили (2.1), мы не имеем права интегрировать по частям, как не раз это делали в § 1. Проблему можно решить, если для перебрасывания производных, хотя бы в некоторых точках, воспользоваться принципом максимума. Ниже это будет продемонстрировано на примере вязкостного решения.

Таким образом, возникает необходимость вводить понятие обобщенного решения и развивать теорию и методы построения этих решений. Такие теории активно создаются и развиваются в течение последних 50-ти лет. Среди получивших признание и стремительно развивающихся в последнее время концепций: энтропийные решения С.Н. Кружкова [10], [45], [53] – [55]; вязкостные решения M.G. Crandall’а и P.L. Lions’а [49], [56]; обобщенные решения на базе идемпотентного анализа [57] – [59]; теория минимаксных решений А.И. Субботина [52], которая берет истоки в теории позиционных дифференциальных игр Н.Н. Красовского [60]; теория сингулярных характеристик А.А. Меликяна [61] – [63].

В работах С.Н. Кружкова начиная с середины 60-ых исследуется в основном случай выпуклого гамильтониана или выпуклой начальной функции. В этих случае есть явные формулы О.А. Олейник, С.Н. Кружкова, E. Hopf'a для решения задачи (2.1), (2.2) [23], [26], [стр. 537 – 557, 9], [45], [55], [главы 3, 10, 49], [стр. 110 – 121, 52], [64] – [66]. Эти формулы содержат операции $\inf$ и $\sup$, отсюда и пошло название "минимаксные решения". Минимаксное представление решений квазилинейных уравнений первого порядка, типа уравнений Гамильтона–Якоби, нашло приложение в 70-ые годы в теории дифференциальных игр (см. работы Н.Н. Красовского, А.И. Субботина, [52], [60] и цитированную там литературу).

Уравнение Г–Я, как уже отмечалось, является вполне нелинейным уравнением, т.е. в общем случае оно не сводится к линейному с помощью удачной замены, подсказанной, например, аппаратом группового анализа [§ 13, 9], [67]. Однако уравнения Г–Я может быть интерпретировано как линейное в идемпотентном полукольце. В конце 80-ых годов коллектив во главе с В.П. Масловым стал разрабатывать аппарат идемпотентного анализа [57], [глава 4, 58]. Было предложено заменить структуру поля над $R^1$ с операциями $x+y, x\cdot y$, структурой полукольца с операциями $x\oplus y=\min\{x,y\}$, $x\otimes y=x+y$. Идемпотентным аналогом интегрирования будет $\inf$, а аналогом преобразования Фурье будет преобразование Фенхеля–Лежандра.

С помощью сингулярных характеристик А.А. Меликяна можно строить некоторые поверхности, на которых решение и (или) гамильтониан негладкие.

Подход А.И. Субботина может рассматриваться, как неклассический метод характеристик [глава 1, 52]. Отметим, что в [глава 2, 52] содержится доказательства эквивалентности различных оп-ределений обобщенного решения начальной задачи для уравнения Г–Я.

В конце 50-ых О.А. Олейник в работе [13] предложила оригинальный подход к исследованию задачи Коши (1.1), (1.2). Этот подход основывается на связи между законом сохранения (1.1) и уравнением Г–Я (2.1). Ниже приводится обобщение этого подхода со случая $\eta(u)=u$ на случай $\eta'(u)>0$.

Согласно результатам § 1 решения следующих уравнений

$$\frac{\partial\eta(u)}{\partial t}+\frac{\partial\varphi(u)}{\partial x}=\varepsilon\frac{\partial^2 u}{\partial x^2}\ ,\ \frac{\partial\eta(u)}{\partial t}+\frac{\partial\varphi(u)}{\partial x}=\varepsilon\frac{\partial^2\eta(u)}{\partial x^2}\quad \text{при } \eta'(u)>0$$

имеет почти всюду одинаковый предел при $\varepsilon\to 0+$, т.е. под обобщенным решением (1.1) мы можем понимать почти всюду $u(t,x)=\lim\limits_{\varepsilon\to 0+}u_\varepsilon(t,x)$, где $u_\varepsilon(t,x)$ – решение уравнения

$$\frac{\partial\eta(u_\varepsilon)}{\partial t}+\frac{\partial\varphi(u_\varepsilon)}{\partial x}=\varepsilon\frac{\partial^2\eta(u_\varepsilon)}{\partial x^2}\ . \qquad (2.3)$$

Введем потенциал $U_\varepsilon(t,x)$ по аналогии с работами [стр. 578 – 580, 9], [13], следующим образом

$$U_\varepsilon(t,x)=\int\limits_{(0,0)}^{(t,x)}\eta\big(u_\varepsilon(t,x)\big)dx+\left[\varepsilon\frac{\partial\eta\big(u_\varepsilon(t,x)\big)}{\partial x}-\varphi\big(u_\varepsilon(t,x)\big)\right]dt. \qquad (2.4)$$

Согласно (2.3) контурный интеграл (2.4) не зависит от пути интегрирования и определяет непрерывную дифференцируемую функцию, производные которой удовлетворяют равенствам

$$\frac{\partial U_\varepsilon}{\partial x}=\eta\big(u_\varepsilon\big),\ \frac{\partial U_\varepsilon}{\partial t}=\varepsilon\frac{\partial\eta\big(u_\varepsilon\big)}{\partial x}-\varphi\big(u_\varepsilon\big). \qquad (2.5)$$

Исключая из (2.5) $\eta\big(u_\varepsilon\big)$, получим

$$\frac{\partial U_\varepsilon}{\partial t}+\varphi\left(\eta^{-1}\left(\frac{\partial U_\varepsilon}{\partial x}\right)\right)=\varepsilon\frac{\partial^2 U_\varepsilon}{\partial x^2}, \qquad (2.6)$$

$$U_\varepsilon(0,x)=U_0(x)=\int\limits_0^x\eta\big(u_0(\xi)\big)d\xi. \qquad (2.7)$$

Из утверждения 1, замечания 1 приложения и того, что начальная функция $u_0(x)$ ограничена, имеем

$$\left|\frac{\partial U_\varepsilon}{\partial x}(t,x)\right| = \left|\eta\left(u_\varepsilon(t,x)\right)\right| \le \max_{x\in R^1}\left|\eta\left(u_0(x)\right)\right| = M_1,$$

$$\left|\frac{\partial U_\varepsilon}{\partial t}(t,x)\right| = \left|\eta'\left(u_\varepsilon(t,x)\right)\left(\varepsilon\frac{\partial u_\varepsilon}{\partial x}(t,x)\right) - \varphi\left(u_\varepsilon(t,x)\right)\right| \le M_2,$$

поэтому

$$|U_\varepsilon(t,x)| \le M_1|x| + M_2 t, \qquad (2.8)$$

где $M_1$, $M_2$ не зависят от $\varepsilon > 0$.

Введём определения.

**Определение 1 (О.А. Олейник).** *Ограниченная равномерно непрерывная функция $U(t,x)$ называется вязкостным решением начальной задачи Коши для уравнения Г–Я (2.1), (2.2), если:*

$$U(t,x) = \lim_{\varepsilon\to 0+} U_\varepsilon(t,x), \qquad (2.9)$$

*где $U_\varepsilon(t,x)$ – решение задачи Коши (2.6), (2.2).*

**Определение 2 (О.А. Олейник).** *Ограниченная измеримая функция $u(t,x)$ называется вязким решением задачи Коши (1.1), (1.2), если:*

1) *Существует вязкостное решение $U(t,x)$ (2.9) задачи Коши (2.1), (2.7);*
2) $u(t,x) = \eta^{-1}\left(\dfrac{\partial U(t,x)}{\partial x}\right)$ *почти всюду по $x \in R^1$ при фиксированном $t \ge 0$.*

Из теорем 1, 2 § 1 и изложенной выше конструкции вязкостных и вязких решений имеем следующее утверждение.

**Утверждение 1.** *Если вязкое решение задачи Коши (1.25), (1.26) совпадает с обобщенным решением (1.25), (1.26), то вязкое решение задачи Коши (1.1), (1.2) совпадает с обобщенным решением (1.1), (1.2).*

Рассмотрим теперь подробнее вязкостные решения M.G. Crandall'a и P.L. Lions'a и сопоставим их с вязкостными решениями О.А. Олейник. Приведенная ниже конструкция базируется на определениях вязкостного решения и минимаксного решения в смысле А.И. Субботина [стр. 15 – 17, 70]. Эта конструкция позволяет достаточно быстро придти к явной формуле E. Hopf'a решения начальной задачи Коши для уравнения Г–Я, при выпуклой начальной функции.

Снова рассмотрим приближенную задачу

$$\frac{\partial U_\varepsilon}{\partial t}+H\left(\frac{\partial U_\varepsilon}{\partial x}\right)=\varepsilon\frac{\partial^2 U_\varepsilon}{\partial x^2}$$

и поставим то же начальное условие (2.2). Подобно тому, как это делалось в § 1, можно показать, что

$$U_\varepsilon \to U \text{ при } \varepsilon \to 0+$$

локально равномерно в $\left(0,+\infty\right)\times R^1$ , причем семейство $\left\{U_\varepsilon\right\}_{\varepsilon>0}$ ограничено и равностепенно непрерывно на компактных подмножествах $\left(0,+\infty\right)\times R^1$ [стр. 464, 49].

Фиксируем какую-либо пробную функцию $V(t,x)\in C^\infty\left(\left(0,+\infty\right)\times R^1\right)$ и предположим, что $U-V$ имеет строгий локальный максимум в некоторой точке $(t_0,x_0)\in\left(0,+\infty\right)\times R^1$ . $U_\varepsilon-V$ имеет локальный максимум в точке $(t_\varepsilon,x_\varepsilon)\in\left(0,+\infty\right)\times R^1$ , причем $(t_\varepsilon,x_\varepsilon)\to(t_0,x_0)$ при $\varepsilon\to 0+$. Тогда,

$$\frac{\partial U_\varepsilon}{\partial x}\left(t_\varepsilon,x_\varepsilon\right)=\frac{\partial V}{\partial x}\left(t_\varepsilon,x_\varepsilon\right),\ \frac{\partial U_\varepsilon}{\partial t}\left(t_\varepsilon,x_\varepsilon\right)=\frac{\partial V}{\partial t}\left(t_\varepsilon,x_\varepsilon\right),$$

$$-\frac{\partial^2 U_\varepsilon}{\partial x^2}\left(t_\varepsilon,x_\varepsilon\right)\ge-\frac{\partial^2 V}{\partial x^2}\left(t_\varepsilon,x_\varepsilon\right).$$

Вычислим

$$\frac{\partial V}{\partial t}\left(t_\varepsilon,x_\varepsilon\right)+H\left(\frac{\partial V}{\partial x}\left(t_\varepsilon,x_\varepsilon\right)\right)=\frac{\partial U_\varepsilon}{\partial t}\left(t_\varepsilon,x_\varepsilon\right)+H\left(\frac{\partial U_\varepsilon}{\partial x}\left(t_\varepsilon,x_\varepsilon\right)\right)=$$

$$= \varepsilon \frac{\partial^2 U_\varepsilon}{\partial x^2}\left(t_\varepsilon, x_\varepsilon\right) \le \varepsilon \frac{\partial^2 V}{\partial x^2}\left(t_\varepsilon, x_\varepsilon\right).$$

Устремив $\varepsilon \to 0+$, используя то, что $(t_\varepsilon, x_\varepsilon) \to (t_0, x_0)$, $V(t,x) \in C^\infty\left(\left(0,+\infty\right)\times R^1\right)$ и $H$ – непрерывная функция, получим

$$\frac{\partial V}{\partial t}\left(t_0, x_0\right) + H\left(\frac{\partial V}{\partial x}\left(t_0, x_0\right)\right) \le 0.$$

Предположение, что $U - V$ имеет строгий локальный минимум в точке $(t_0, x_0) \in \left(0,+\infty\right)\times R^1$ можно ослабить, убрав требование строгости. Действительно, вместо $V(t,x)$ следует рассмотреть

$$\tilde{V}(t,x) = V(t,x) + \delta\left(\left(x - x_0\right)^2 + \left(t - t_0\right)^2\right),\ \delta > 0.$$

Аналогично рассматривается случай, когда $U - V$ имеет локальный максимум. Таким образом, мы пришли к следующему определению [56], эквивалентному определению 1.

**Определение 3 (P.L. Lions, M.G. Crandall).** *Ограниченная равномерно непрерывная функция $U(t,x)$ называется вязкостным решением начальной задачи Коши для уравнения Г–Я (2.1), (2.2), если:*

*1) $U\left(0,x\right) = U_0(x)$ , $x \in R^1$;*

*2) для каждой $V(t,x) \in C^\infty\left(\left(0,+\infty\right)\times R^1\right)$ выполняются следующие условия:*

*если $U - V$ имеет локальный максимум в точке $(t_0, x_0) \in \left(0,+\infty\right)\times R^1$ , то*

$$\frac{\partial V}{\partial t}\left(t_0, x_0\right) + H\left(\frac{\partial V}{\partial x}\left(t_0, x_0\right)\right) \le 0 \qquad (2.10),$$

*если $U - V$ имеет локальный минимум в точке $(t_0, x_0) \in \left(0,+\infty\right)\times R^1$ , то*

$$\frac{\partial V}{\partial t}\left(t_0, x_0\right) + H\left(\frac{\partial V}{\partial x}\left(t_0, x_0\right)\right) \ge 0. \qquad (2.11)$$

Заметим, что в [стр. 70 – 71, 52] знаки в неравенствах (2.10), (2.11) противоположные. Пункт 1 определения 3 и (2.10) определяет субрешение (нижнее решение), а (2.11) – суперрешение (верхнее решение). Таким образом, вязкостное решение – это одновременно субрешение и суперрешение. Имеет место следующий результат, доказательство которого базируется на принципе максимума [стр. 463 – 472, 49].

**Теорема 1.** *Вязкостное решение* $U(t,x)$ *существует и притом единственно. Кроме того, если* $U(t,x)$ *дифференцируема в некоторой точке* $(t_0, x_0)$*, то*

$$\frac{\partial U}{\partial t}(t_0,x_0)+H\left(\frac{\partial U}{\partial x}(t_0,x_0)\right)=0.$$

Покажем, где используются вязкостные решения. Для этого рассмотрим следующую задачу оптимального управления [глава 4, 71]:

$$J\left(\tau,x;\,u(\cdot)\right)=\int_{t_0}^{\tau} L\left(t,x(t),u(t)\right)dt+\varphi(t_0,x_0),\ \dot{x}(t)=f\left(t,x(t),u(t)\right),$$

$$u(t)\in \mathrm{M},\ x(\tau)=x.$$

Положим функцию цены $U(\tau,x)$ равной

$$U(\tau,x)=\inf_{u(\cdot)\in \mathrm{M}} J\left(\tau,x;\,u(\cdot)\right),\ U(t_0,x)=\varphi(t_0,x).$$

Тогда справедлива следующая теорема [стр. 472 – 481, 49].

**Теорема 2.** $U(\tau,x)$ *является вязкостным решением начальной задачи Коши для уравнения Гамильтона–Якоби–Беллмана*[8]

$$\frac{\partial U}{\partial t}+\sup_{u\in \mathrm{M}}\left\{\left\langle\frac{\partial U}{\partial x},\,f\left(t,x,u\right)\right\rangle-L\left(t,x,u\right)\right\}=\frac{\partial U}{\partial t}+H\left(\frac{\partial U}{\partial x}\right)=0,$$

$$U(t_0,x)=\varphi(t_0,x).$$

Следуя А.И. Субботину [стр. 292, 52] введем определение.

[8] Это уравнение также называют прямым уравнение Г–Я–Б [72]. В литературе по динамическому программированию чаще встречается попятное уравнение Г–Я–Б.

**Определение 4.** *Верхней и нижней производной по Дини (или Хаддаду) функции* $U(t,x)$ *в точке* $(t_0,x_0)$ *по направлению* $\vec{f}=(f_1,f_2)$ *называются соответственно:*

$$d^+U\left(t_0,x_0;\vec{f}\right)=$$
$$=\lim_{\varepsilon\downarrow 0}\sup\left\{\frac{U\left(t_0+\delta f_1',x_0+\delta f_2'\right)-U\left(t_0,x_0\right)}{\delta}:\left(\delta,\vec{f}'\right)\in\Delta_\varepsilon\left(t_0,x_0;\vec{f}\right)\right\},$$
$$d^-U\left(t_0,x_0;\vec{f}\right)=$$
$$=\lim_{\varepsilon\downarrow 0}\inf\left\{\frac{U\left(t_0+\delta f_1',x_0+\delta f_2'\right)-U\left(t_0,x_0\right)}{\delta}:\left(\delta,\vec{f}'\right)\in\Delta_\varepsilon\left(t_0,x_0;\vec{f}\right)\right\},$$

*где* $\Delta_\varepsilon\left(t_0,x_0;\vec{f}\right)=\left\{\left(\delta,\vec{f}'\right)\in(0,\varepsilon)\times R^2:\left\|\vec{f}-\vec{f}'\right\|\le\varepsilon\right\}$.

*Супердифференциалом и субдифференциалом функции* $U(t,x)$ *в точке* $(t_0,x_0)$ *будем называть соответственно следующие множества:*

$$D^+U\left(t_0,x_0\right)=\left\{\vec{s}\in R^2:\left\langle\vec{s},\vec{f}\right\rangle-d^+U\left(t_0,x_0;\vec{f}\right)\ge 0,\ \forall\,\vec{f}\right\},\qquad(2.12)$$
$$D^-U\left(t_0,x_0\right)=\left\{\vec{s}\in R^2:\left\langle\vec{s},\vec{f}\right\rangle-d^-U\left(t_0,x_0;\vec{f}\right)\le 0,\ \forall\,\vec{f}\right\}.\qquad(2.13)$$

Понимая под $U\left(t_0,x_0\right)$ – вязкостное решение, мы можем считать, что $d^+U\left(t_0,x_0;\vec{f}\right)=d^-U\left(t_0,x_0;\vec{f}\right)=dU\left(t_0,x_0;\vec{f}\right)$, т.е. $U\left(t_0,x_0\right)$ имеет производную по любому направлению $\vec{f}$ в любой $(t_0,x_0)\in(0,+\infty)\times R^1$ [стр. 112, 295 – 297, 52].

Заметим, что $D^+U\left(t_0,x_0\right)$, $D^-U\left(t_0,x_0\right)$ – замкнутые выпуклые множества, которые могут быть и пустыми. Если $U\left(t,x\right)$ дифференцируема в точке $\left(t_0,x_0\right)$, то

$$D^+U\left(t_0,x_0\right)=D^-U\left(t_0,x_0\right)=\left\{DU\left(t_0,x_0\right)\right\},$$

где $DU(t_0,x_0)$ – градиент функции $U(t,x)$ в точке $(t_0,x_0)$. Так же важно отметить, что $D^{-}U(t_0,x_0)$ – это обобщение классического субдифференциала $\partial U(t_0,x_0)$ из выпуклого анализа [стр. 42 – 52, 73]. Легко привести пример не выпуклой гладкой функции $U(t,x)$, для которой $D^{-}U(t_0,x_0)=\{DU(t_0,x_0)\}$, а $\partial U(t_0,x_0)=\{\varnothing\}$.

Пусть $V(t,x)\in C^{\infty}\left((0,+\infty)\times R^1\right)$ и выполняется следующее условие: $U-V$ имеет локальный максимум в точке $(t_0,x_0)\in(0,+\infty)\times R^1$. В таком случае, $D^{-}U(t_0,x_0)\neq\{\varnothing\}$,

$$\forall\ \vec{f}\to dU\left(t_0,x_0;\vec{f}\right)\le dV\left(t_0,x_0;\vec{f}\right),$$

$$\left\langle\left(\frac{\partial V}{\partial t},\frac{\partial V}{\partial x}\right),\vec{f}\right\rangle=dV\left(t_0,x_0;\vec{f}\right)\ge dU\left(t_0,x_0;\vec{f}\right),\text{ то}$$

$$\left(\frac{\partial V}{\partial t},\frac{\partial V}{\partial x}\right)\in D^{+}U(t_0,x_0).$$

Легко видеть, что и обратно, если $(s_1,s_2)\in D^{-}U(t_0,x_0)\neq\{\varnothing\}$, то $\exists V(t,x)\in C^{\infty}\left((0,+\infty)\times R^1\right)$: $\left(\frac{\partial V}{\partial t}(t_0,x_0),\frac{\partial V}{\partial x}(t_0,x_0)\right)=(s_1,s_2)$ и $U-V$ имеет локальный минимум в точке $(t_0,x_0)\in(0,+\infty)\times R^1$. Аналогичные рассуждения можно провести и для суперрешения.

Введем индикаторную функцию

$$I_A(\vec{x})=\begin{cases}0, & \vec{x}\in A\\ +\infty, & \vec{x}\notin A\end{cases}$$

и обозначим $F(\vec{s})=s_1+H(s_2)$.

С помощью определения 4 можно записать определение 3 в более конструктивном виде.

**Определение 5.** *Ограниченная равномерно непрерывная функция* $U(t,x)$ *называется вязкостным решением начальной задачи Коши для уравнения Г–Я (2.1), (2.2), если:*

1) $U(0,x)=U_0(x)$ , $x\in R^1$;

2) $\forall\,(t,x)\in(0,+\infty)\times R^1,\ \forall\,\vec{s}\in R^2$

$$F(\vec{s})\geq -I_{D^-U(t,x)}(\vec{s}) \text{ – суперрешение,} \tag{2.14}$$

$$F(\vec{s})\leq I_{D^+U(t,x)}(\vec{s}) \text{ – субрешение.} \tag{2.15}$$

Из (2.12), (2.13) следует, что

$$I_{D^-U(t,x)}(\vec{s})=\sup_{\vec{f}\in R^2}\left\{\left\langle\vec{s},\vec{f}\right\rangle-dU\left(t_0,x_0;\vec{f}\right)\right\},$$

$$I_{D^+U(t,x)}(\vec{s})=\sup_{\vec{f}\in R^2}\left\{-\left\langle\vec{s},\vec{f}\right\rangle+dU\left(t_0,x_0;\vec{f}\right)\right\}. \tag{2.16}$$

Используя (2.16), (2.14) и (2.15) можно переписать следующим образом [стр. 57, 70]:

$$\forall\,(t,x)\in(0,+\infty)\times R^1,\ \forall\,\vec{s}\in R^2$$

$$\inf_{\vec{f}\in R^2}\left\{dU\left(t_0,x_0;\vec{f}\right)-\left\langle\vec{s},\vec{f}\right\rangle-F(\vec{s})\right\}\leq 0 \text{ – суперрешение,} \tag{2.17}$$

$$\sup_{\vec{f}\in R^2}\left\{dU\left(t_0,x_0;\vec{f}\right)-\left\langle\vec{s},\vec{f}\right\rangle-F(\vec{s})\right\}\geq 0 \text{ – субрешение.} \tag{2.18}$$

Интересно заметить, что минимаксное решение, которое эквивалентно вязкостному, определяется А.И. Субботиным в [стр. 70 – 72, 52], как:

$$\forall\,(t,x)\in(0,+\infty)\times R^1,\ \forall\,\vec{s}\in R^2$$

$$\inf_{\vec{f}\in R^2}\left\{dU\left(t_0,x_0;\vec{f}\right)-\left\langle\vec{s},\vec{f}\right\rangle+F(\vec{s})\right\}\leq 0 \text{ – верхнее,}$$

$$\sup_{\vec{f}\in R^2}\left\{dU\left(t_0,x_0;\vec{f}\right)-\left\langle\vec{s},\vec{f}\right\rangle+F(\vec{s})\right\}\geq 0 \text{ – нижнее.}$$

Следующая теорема о дифференцировании под знаком супремума [стр. 295 – 297, 52], [§ 3, § 10, 74], поможет нам доказать впоследствии теорему E. Hopf'a.

**Теорема 3 (Демьянов–Данскин).** *Пусть*

$$U(t,x)=\sup_{s\in R^1}\varphi(t,x,s)$$

*и выполнены условия*

1) $\varphi(t,x,s)=\varphi_1(t,x,s)-\varphi_2(s)$;

2) $\varphi_1(t,x,s)$ – *непрерывна; при любом* $s\in R^1$ $\varphi_1(\cdot,s)$ – *дифференцируема, причем функция* $(t,x,s)\to D_{t,x}\varphi_1(t,x,s)$ *непрерывна;*

3) $\exists a:\varphi_2(s)\ge a$; $\varphi_2(s)$ – *полунепрерывная снизу функция; эффективная область* $\varphi_2(s)$ *является непустым ограниченным множеством, т.е.*

$$dom\ \varphi_2=\{s\in R^1:\varphi_2(s)<+\infty\}.$$

*Тогда* $U(t,x)$ – *липшицева функция в каждой ограниченной выпуклой области. В каждой точке* $U(t,x)$ *дифференцируема по любому направлению. Более того,*

$$d^+U\left(t_0,x_0;\vec f\right)=d^-U\left(t_0,x_0;\vec f\right)=dU\left(t_0,x_0;\vec f\right)=$$

$$=\max_{\bar s\in S_0(t,x)}\left\langle D_{t,x}\varphi_1(t,x,s),\vec f\right\rangle=\max_{\bar s\in S_0(t,x)}\left\{\frac{\partial\varphi_1(t,x,s)}{\partial t}f_1+\frac{\partial\varphi_1(t,x,s)}{\partial x}f_2\right\},$$

*где* $S_0(t,x)=Arg\max_{s\in R^1}\varphi(t,x,s)$ – *непустое ограниченное замкнутое множество, в силу условий 1) – 3).*

Теперь приведем теорему доказанную E. Hopf'ом [65] для минимаксных решений, а позже M. Bardi и L.C. Evans'ом [66] и для вязкостных.

**Теорема 4 (E. Hopf).** *Если* $U_0(x)$ – *выпуклая функция, то вязкостное решение (2.1), (2.2) имеет вид*

$$U(t,x)=\sup_{s\in R^1}\left[\langle s,x\rangle-H(s)t-U_0^*(s)\right],\tag{2.19}$$

*где* $U_0^*(s)=\sup_{x\in R^1}\left[\langle s,x\rangle-U_0(x)\right]$;

*Если* $H(s)$ – *выпуклая функция, то вязкостное решение (2.1), (2.2) имеет вид*

$$U(t,x)=\sup_{f\in R^1}\left[U_0(x-tf)+H^*(f)t\right],\tag{2.20}$$

*где* $H^*(f) = \sup\limits_{s \in R^1} \left[ \langle s, f \rangle - H(s) \right]$.

**Доказательство.** Докажем для определенности первое утверждение (2.19). Доказательство (2.20) можно провести похожим образом.

Выполнение начального условия (пункт 1 в определении вязкостного решения), следует из теоремы Фенхеля–Моро о том, что "вторая сопряженная к собственной замкнутой выпуклой функции совпадает с ней самой" [стр. 41, 73]. Сопряженная функция $U_0^*(s) = \sup\limits_{x \in R^1} \left[ \langle s, x \rangle - U_0(x) \right]$ является полунепрерывной снизу, как верхняя грань семейства аффинных функций, и удовлетворяет оценке $U_0^*(s) \geq a = -U_0^*(0)$. Предположим сначала, что множество $dom\, U_0^*(s) = \left\{ s \in R^1 : U_0^*(s) < +\infty \right\}$ – ограничено. Это предположение справедливо, если $U_0(x)$ равномерно на $R^1$ удовлетворяет условию Липшица[9] [стр. 134, 75]. Нетрудно доказать, что в этом случае множество

$$S_0(t,x) = \left\{ s_0 \in R^1 : \langle s_0, x \rangle - H(s_0)t - U_0^*(s_0) = U(t,x) \right\}$$

является непустым компактом в $R^1$.

По теореме 3 имеем, что

$$dU\left(t,x;\vec{f}\right) = \max_{s_0 \in S_0(t,x)} \left\{ -H(s_0) f_1 + s_0 f_2 \right\}.$$

Подставив выражение для $dU\left(t,x;\vec{f}\right)$ в (2.2.17). (2.2.18), получим

$$\forall\, (t,x) \in (0,+\infty) \times R^1,\ \forall\, \vec{s} \in R^2 \rightarrow$$

$$\inf_{\vec{f} \in R^2} \max_{s_0 \in S_0(t,x)} \left\{ -H(s_0) f_1 + s_0 f_2 - \left\langle \vec{s}, \vec{f} \right\rangle - F(\vec{s}) \right\} \leq 0, \qquad (2.21)$$

$$\forall\, (t,x) \in (0,+\infty) \times R^1,\ \forall\, \vec{s} \in R^2 \rightarrow$$

---

[9] В случае начальных данных (2.7) это условие, очевидно, выполняется. Более того, в силу оценки (2.8) $U(t,x)$ для любого момента времени $t$ равномерно на $R^1$ удовлетворяет условию Липшица.

$$\sup_{\vec{f}\in R^2}\max_{s_0\in S_0(t,x)}\left\{-H(s_0)f_1+s_0f_2-\left\langle\vec{s},\vec{f}\right\rangle-F(\vec{s})\right\}\geq 0. \qquad (2.22)$$

Перепишем (2.21) следующим образом:

$$\forall(s_1,s_2)\in R^2\rightarrow$$

$$\inf_{(f_1,f_2)}\max_{s_0\in S_0(t,x)}\left\{-H(s_0)f_1+s_0f_2-s_1f_1-s_2f_2-s_1-H(s_2)\right\}=$$

$$=\inf_{(f_1,f_2)}\max_{s_0\in S_0(t,x)}\left\{-(s_1+H(s_0))f_1+f_2(s_0-s_2)-s_1-H(s_2)\right\}\leq 0. \qquad (2.23)$$

По теореме о минимаксе [стр. 680 – 681, 25], inf и max в (2.23) можно поменять местами [стр. 112 – 115, 70].

Так как $(f_1,f_2)$ – могут принимать любые значения, а множество $S_0(t,x)$ – ограниченное, то если не выполнено хотя бы одно из следующих равенств:

$$s_1+H(s_0)=0,\ s_0-s_2=0, \qquad (2.24)$$

то $\max_{s_0\in S_0(t,x)}\inf_{(f_1,f_2)}\left\{-(s_1+H(s_0))f_1+f_2(s_0-s_2)-s_1-H(s_2)\right\}=-\infty$ и (2.21) выполняется. Поэтому, можно считать, что (2.24) выполняется. Тогда

$$-(s_1+H(s_0))f_1+f_2(s_0-s_2)-s_1-H(s_2)=$$

$$=-s_1-H(s_2)=-(s_1+H(s_0))=0.$$

Таким образом, доказана справедливость (2.21).

Аналогично докажем (2.22). Если выполняется (2.24), то

$$\sup_{(f_1,f_2)}\max_{s_0\in S_0(t,x)}\left\{-(s_1+H(s_0))f_1+f_2(s_0-s_2)-s_1-H(s_2)\right\}=0,$$

иначе

$$\sup_{(f_1,f_2)}\max_{s_0\in S_0(t,x)}\left\{-(s_1+H(s_0))f_1+f_2(s_0-s_2)-s_1-H(s_2)\right\}=+\infty.$$

Таким образом, доказана справедливость (2.23).

Покажем, что функция $U(t,x)$ вида (2.19) является вязкостным решением и в общем случае, без дополнительной гипотезы о начальной функции $U_0(x)$. Введем следующие обозначения

$$U_0^k(x) = \max_{|s|\le k}\left[\langle s,x\rangle - U_0^*(s)\right],$$

$$U^k(t,x) = \max_{|s|\le k}\left[\langle s,x\rangle - H(s)t - U_0^*(s)\right],\ k\in N.$$

Из доказательства проведенного выше следует, что $U^k(t,x)$ является вязкостным решением начальной задачи Коши для уравнения Г–Я

$$\frac{\partial U^k}{\partial t} + H\left(\frac{\partial U^k}{\partial x}\right) = 0\ ,\ U^k(0,x) = U_0^k(x)\ .$$

Последовательность $U^k(t,x)$ – возрастающая, поэтому существует предел

$$\lim_{k\to\infty} U^k(t,x) = U(t,x).$$

Нетрудно проверить [стр. 112 – 115, 70], что предельная функция и есть вязкостное решение. Легко показать, что предельная функция $U(t,x)$ удовлетворяет начальному условию (2.2). С другой стороны, формула (2.19) справедлива и для предельной функции.

Теорема 4 доказана.

**Замечание.** На данный момент известны различные обобщения этих формул, например, на случай гамильтониана зависящего от времени [76].

Следующее определение приводится для случая, когда $\varphi(\eta^{-1}(y))$ – строго выпуклая функция, т.е. $(\varphi'(u)/\eta'(u))' > 0$. Для случая $\eta(u) = u$ оно приведено в [стр. 126 – 129, 49].

**Определение 6 (P.D. Lax, O.A. Олейник).** *Ограниченную измеримую функцию $u(t,x)$ будем называть обобщенным по Lax'у–Олейник решением задачи Коши (1.1), (1.2) в $\pi_\infty$, если почти всюду по $x$ при фиксированном $t > 0$ выполняется*

$$u(t,x)=\eta^{-1}\left(G\left(\frac{x-\arg\min\limits_{y}\left\{tL\left(\frac{x-y}{t}\right)+\int\limits_{0}^{y}u_0(\xi)d\xi\right\}}{t}\right)\right), \textit{ где} \quad (2.25)$$

$G(x)=\left(\left(\varphi\left(\eta^{-1}(\cdot)\right)\right)'\right)^{-1}(x)=\left(\frac{\varphi'(\cdot)}{\eta'(\cdot)}\right)^{-1}(x)$ – *обратная функция к* $\left(\frac{\varphi'(y)}{\eta'(y)}\right)$; $L(x)=\sup\limits_{p}\left\{px-\varphi\left(\eta^{-1}(p)\right)\right\}$ – *функция, сопряжена по Фенхелю–Лежандру к* $\varphi\left(\eta^{-1}(p)\right)$.

Эквивалентность этого определения при $H''=(\varphi'/\eta')'>0$ определению 1 § 1 следует из работ [стр. 578 – 580, 9], [13] и утверждения 1.

В работах Н.Н. Кузнецова и Б.Л. Рождественского [стр. 537 – 557, 9], [68], [69] предлагается альтернативный по сравнению с [23], [26] способ получения явных формул для решения (1.1), (1.2) при $\eta(u)=u$ и строго выпуклой $\varphi(u)$. Этот метод базируется на исследовании характеристической системы для потенциала (2.4) $\Phi(t,x)=U_\varepsilon(t,x)\big|_{\varepsilon=0}$ [68], [стр. 12 – 16, 52]

$$\frac{dx}{dt}=H'_u,\quad \frac{du}{dt}=0,\quad \frac{d\Phi}{dt}=uH'_u-H. \qquad (2.26)$$

В случае строго выпуклой функции $\varphi\left(\eta^{-1}(u)\right)$ характеристики, стартующие в момент времени $t=0$ из разных точек $x_0^1<x_0^2$, приходят в момент времени $t>0$ в одну и ту же точку $x$, только если $u\left(t,x;0,x_0^1,u_0^1\in u_0\left(x_0^1\right)\right)=u_0^1>u_0^2=u\left(t,x;0,x_0^2,u_0^2\in u_0\left(x_0^2\right)\right)$. Исходя из этого, была получена формула для решения $\Phi(t,x)$ [68], [69]. Обозначим $\Phi(t,x;x_0,u_0)$ решение третьего уравнения системы (2.26) при начальных условиях в момент времени $t=0$:

$$x(0)=x_0\,,\; u(0)=u_0\in u_0(x_0)\,,\; \Phi(0)=\Phi_0=\int_0^{x_0}\eta\big(u_0(\xi)\big)d\xi\,.$$

Тогда

$$\Phi(t,x)=\min_{x_0,u_0\in u_0(x_0)}\Phi(t,x;x_0,u_0) \qquad (2.27)$$

является непрерывной по $t$, $x$ функцией, удовлетворяющей почти всюду (2.1), а $u(t,x)=\eta^{-1}\left(\dfrac{\partial\Phi(t,x)}{\partial x}\right)$ является обобщенным решением (1.1). Формула (2.27) допускает дальнейшие упрощения, конечным итогом которых будет формула Lax'а–Олейник (2.25).

Следующее определение приводится для случая, когда $u_0(x)$ – монотонная функция (для определенности возрастающая), имеющая пределы слева $u_-$ и справа $u_+$.

**Определение 7 (E. Hopf, С.Н. Кружков, Н.С. Петросян).** *Ограниченная измеримая функция* $u(t,x)$ *называется обобщенным по E. Hopf'у решением задачи (1.1), (1.2) в слое* $\pi_\infty$, *если*[10]:

$$u(t,x)=\eta^{-1}\left(\arg\sup_{s\in(u_-,u_+)}\left\{-\sup_y\left[\int_0^y\big(s-u_0(x)\big)dx\right]+sx-t\varphi\big(\eta^{-1}(s)\big)\right\}\right).$$

Поясним происхождение этого определения. Можно показать (аналог теоремы Демьянова–Данскина), что для $\eta(u)=u$ [45]

$$u(t,x)=\frac{\partial U}{\partial x}(t,x)\in S_0(t,x)=$$

$$=\left\{s_0\in R^1:\langle s_0,x\rangle-H(s_0)t-U_0^*(s_0)=U(t,x)\right\}.$$

[10] Следует заметить, что супремум может достигаться не в одной точке, однако, оказывается, что все селекторы многозначного отображения $\operatorname{Arg}\sup\limits_{p\in(u_-,u_+)}\{...\}$ совпадают почти всюду по $x$ при фиксированном $t\geq 0$. Заметим, что обобщенное решение задачи Коши (1.1), (1.2) также определяется с точностью до почти всюду по $x$ при фиксированном $t\geq 0$.

В той же работе [45] показано, что при $\eta(u)=u$ $u(t,x)=\frac{\partial U}{\partial x}(t,x)$ – есть обобщенное решение (1.1), (1.2) в смысле определения 3 § 1, и что множество $S_0(t,x)$ почти при всех $x$ при фиксированном $t>0$ состоит из одной точки, и тем $x$, в которых $S_0(t,x)$ содержит более одной точки, соответствует разрыв решения при данном $t>0$. Из утверждения 1 следует, что определение 7 эквивалентно определению 3 § 1 при $\eta'(u)>0$.

Приводимое ниже утверждение является обобщением теоремы 1 § 1.

**Утверждение 2.** *Определениям 1 – 8 § 1 и 6, 7 § 2 удовлетворяет одна и та же единственная (с точностью до почти всюду по* $x$ *при фиксированном* $t>0$*) функция* $u(t,x)$, *которую будем называть обобщенным решением. Кроме того, имеет место непрерывная зависимость* $u(t,x)$ *от* $u_0(x)$:

$$\text{если } u_0^k(x)\xrightarrow{L_1}u_0(x), \text{ то } u^k(t,x)\xrightarrow{L_1}u(t,x).$$

## § 3 ДИФФЕРЕНЦИАЛЬНО-РАЗНОСТНЫЙ АНАЛОГ КВАЗИЛИНЕЙНОГО УРАВНЕНИЯ ТИПА ЗАКОНА СОХРАНЕИЯ

Рассмотрим дифференциально-разностное уравнение распределения предприятий по уровням эффективности [77]

$$h\frac{du^n}{dt}=-\Phi(u^n)(u^n-u^{n-1})+\mu\cdot(u^{n+1}-u^n),\; u^n(0)=u_0(nh). \qquad (3.1)$$

Уровни эффективности – $n\in Z$; $u^n(t)\in[0,1]$ – доля предприятий, которые находятся в момент времени $t\geq 0$ на уровнях с номерами не большими, чем $n$; $\mu\geq 0$ – слагаемое характеризующее выбытие мощностей. Если $\Phi(u)=\alpha+\beta\cdot(1-u)$, то говорят, что

$\alpha \geq 0$ – инновационная составляющая, $\beta \geq 0$ – имитационная составляющая. Будем считать, что

$$\Phi(u) > 0\,,\ \ \eta'(u) = \frac{1}{\Phi(u)+\mu}\,,\ \ \varphi'(u) = \frac{\Phi(u)-\mu}{\Phi(u)+\mu}\,,\ \ u_- = 0\,,\ \ u_+ = 1\,,$$

$u_0(x)$ – не убывающая функция, как функция распределения. Тогда

$$\left(\frac{\varphi'(u)}{\eta'(u)}\right)' = \left(\Phi(u)-\mu\right)' = \Phi'(u)\,.$$

Это означает, что $\varphi\left(\eta^{-1}\left(y\right)\right)$ – (строго) выпуклая функция тогда и только тогда, когда $\Phi(u)$ – (строго) монотонно возрастающая функция. Аналогично, (строгая) вогнутость $\varphi\left(\eta^{-1}\left(y\right)\right)$ равносильна (строгому) монотонному убыванию $\Phi(u)$.

Целью настоящего параграфа является установление связи между задачей (3.1) и задачей (1.1), (1.2). Будет показано, что при определенных условиях, значение $u^n\left(t\right)$ в некотором смысле близко к $u\left(t,nh\right)$, где $u\left(t,x\right)$ – решение (1.1), (1.2). Благодаря определению 7 § 2 мы получим приближенную, но явную формулу E. Hopf'а для решения задачи (3.1).

Рассмотрим следующее квазилинейное уравнение типа закона сохранения с вязкостью, которое является обобщением уравнения Бюргерса [глава 4, 1],

$$\frac{\partial u_\varepsilon}{\partial t} + \frac{\varphi'(u_\varepsilon)}{\eta'(u_\varepsilon)}\frac{\partial u_\varepsilon}{\partial x} = \frac{\varepsilon}{\eta'(u_\varepsilon)}\frac{\partial^2 u_\varepsilon}{\partial x^2}\,,\ \ \varepsilon = h/2\,,\ u_\varepsilon(0,x) = u_0(x)\,. \qquad (3.2)$$

Подставим решение этого уравнения $u_\varepsilon(t,x)$ в (3.1), получим (везде в дальнейшем для краткости будем опускать у функций аргумент $t$)

$$\frac{\partial u_\varepsilon(nh)}{\partial t} = -\Phi\left(u_\varepsilon(nh)\right)\frac{\left(u_\varepsilon\left(nh\right)-u_\varepsilon\left((n-1)h\right)\right)}{h} +$$

$$+\mu\frac{\left(u_\varepsilon\left((n+1)h\right)-u_\varepsilon\left(nh\right)\right)}{h}+F(t,nh). \qquad (3.3)$$

Найдем $F(t,nh)$:

$$\frac{\left(u_\varepsilon\left(nh\right)-u_\varepsilon\left((n-1)h\right)\right)}{h}=\frac{\partial u_\varepsilon\left(nh\right)}{\partial x}-\frac{h}{2}\frac{\partial^2 u_\varepsilon\left((n+\theta_1)h\right)}{\partial x^2};$$

$$\frac{\left(u_\varepsilon\left((n+1)h\right)-u_\varepsilon\left(nh\right)\right)}{h}=\frac{\partial u_\varepsilon\left(nh\right)}{\partial x}+\frac{h}{2}\frac{\partial^2 u_\varepsilon\left((n+\theta_2)h\right)}{\partial x^2};$$

$$\frac{\partial u_\varepsilon(nh)}{\partial t}+\left(\Phi\left(u_\varepsilon(nh)\right)-\mu\right)\frac{\partial u_\varepsilon(nh)}{\partial x}=$$

$$=F(t,nh)+\frac{h}{2}\left(\Phi\left(u_\varepsilon(nh)\right)\frac{\partial^2 u_\varepsilon\left((n+\theta_1)h\right)}{\partial x^2}+\mu\frac{\partial^2 u_\varepsilon\left((n+\theta_2)h\right)}{\partial x^2}\right),$$

где $|\theta_1|\le 1$, $|\theta_2|\le 1$. Таким образом,

$$F(t,nh)=\frac{h}{2}\left(\left(\Phi\left(u_\varepsilon(nh)\right)+\mu\right)\frac{\partial^2 u_\varepsilon\left(nh\right)}{\partial x^2}-\right.$$

$$\left.-\Phi\left(u_\varepsilon(nh)\right)\frac{\partial^2 u_\varepsilon\left((n+\theta_1)h\right)}{\partial x^2}-\mu\frac{\partial^2 u_\varepsilon\left((n+\theta_2)h\right)}{\partial x^2}\right). \qquad (3.4)$$

Нам интересно знать, как ведет себя разность $w^n=u^n-u_\varepsilon\left(nh\right)$ при $h\to 0+$, причем $w^n(0)=0$, $n\in N$ – по условию. Поэтому, вычтем из уравнения (3.1) уравнение (3.3). Получим

$$\frac{dw^n}{dt}=-\Phi\left(u^n\right)\frac{\left(w^n-w^{n-1}\right)}{h}+\mu\frac{\left(w^{n+1}-w^n\right)}{h}+$$

$$+\left(\Phi\left(u_\varepsilon(nh)\right)-\Phi\left(u^n\right)\right)\frac{\left(u_\varepsilon\left(nh\right)-u_\varepsilon\left((n-1)h\right)\right)}{h}-F(t,nh)=$$

$$=-\Phi\left(u^n\right)\frac{\left(w^n-w^{n-1}\right)}{h}+\mu\frac{\left(w^{n+1}-w^n\right)}{h}-$$

$$-\left[\left(\int_0^1 \Phi'\left(u^n + t\left(u_\varepsilon(nh) - u^n\right)\right)dt\right)\frac{\partial u_\varepsilon\left((n+\theta)h\right)}{\partial x}\right]w^n - F(t,nh),$$

где $|\theta| \le 1$. Итак,

$$\frac{dw^n}{dt} = -\Phi\left(u^n\right)\frac{\left(w^n - w^{n-1}\right)}{h} + \mu\frac{\left(w^{n+1} - w^n\right)}{h} - C(t,nh)w^n - F(t,nh), \quad (3.5)$$

где $C(t,nh) = \left[\left(\int_0^1 \Phi'\left(u^n + t\left(u_\varepsilon(nh) - u^n\right)\right)dt\right)\frac{\partial u_\varepsilon\left((n+\theta)h\right)}{\partial x}\right]$, $|\theta| \le 1$.

Положим,

$$\Pi = \left\{(t,x) : 0 \le t \le T, \Gamma_-(t) \le x \le \Gamma_+(t)\right\},$$

где $\Gamma_-(0) \le \Gamma_+(0)$ и $\Gamma_-(t)$ – выпуклая функция на $0 \le t \le T$, а $\Gamma_+(t)$ – вогнутая функция на $0 \le t \le T$. Обозначим через $\Gamma$ параболическую границу $\Pi$.

**Лемма (Обобщенный дифференциально-разностный принцип максимума).** *Пусть*

$$\frac{dw^n}{dt} = -\Phi\left(u^n\right)\frac{\left(w^n - w^{n-1}\right)}{h} + \mu\frac{\left(w^{n+1} - w^n\right)}{h} - C(t,nh)w^n - F(t,nh),$$

*и* $\exists\, c_0 : \forall\, (t,nh) \in \Pi \to C(t,nh) \ge c_0$.

*Тогда* $\forall\, \alpha > 0 : \alpha + c_0 > 0$ *выполняется:*

$$\left|w^n\right| \le \max\left\{\max_{(t,nh)\in\Pi}\frac{e^{\alpha T}\left|F(t,nh)\right|}{\alpha + c_0},\ e^{\alpha T}\left|w^n\right|_\Gamma\right\}. \quad (3.6)$$

**Доказательство.** Положим $w^n = e^{\alpha t} v^n$. Тогда (3.5) можно переписать следующим образом:

$$\frac{F(t,nh)}{e^{\alpha t}} + \left(C(t,nh) + \alpha\right)v^n = -\Phi\left(u^n\right)\frac{\left(v^n - v^{n-1}\right)}{h} + \mu\frac{\left(v^{n+1} - v^n\right)}{h} - \frac{dv^n}{dt}.$$

Пусть $v^n$ принимает максимальное значение в некоторой точке $(t,nh) \in \Pi \setminus \Gamma$, не принадлежащей параболической границе, тогда в этой точке либо $v^n \le 0$, либо $v^n \ge 0$ и

$$\frac{F(t,nh)}{\mathrm{e}^{\alpha t}} v^n + \left(C(t,nh)+\alpha\right)\left(v^n\right)^2 =$$

$$= \left( -\Phi\left(u^n\right) \frac{\left(v^n - v^{n-1}\right)}{h} + \mu \frac{\left(v^{n+1} - v^n\right)}{h} - \frac{dv^n}{dt} \right) v^n \le 0,$$

т.к. $v^n \ge 0$, $v^n \ge v^{n-1}$, $v^{n+1} \le v^n$, $\dfrac{dv^n}{dt} \ge 0$. Последнее неравенство может быть строгим, только если $t = T$. Итак, мы получили, что

$$v^n \le \max_{(t,nh)\in\Pi} \frac{\left|F(t,nh)\right|}{\mathrm{e}^{\alpha t}\left(\alpha + c_0\right)} \le \max_{(t,nh)\in\Pi} \frac{\left|F(t,nh)\right|}{\left(\alpha + c_0\right)}.$$

Пусть $v^n$ принимает минимальное значение в некоторой точке $(t,nh) \in \Pi \setminus \Gamma$, не принадлежащей параболической границе, тогда в этой точке либо $v^n \ge 0$, либо $v^n \le 0$ и

$$\frac{F(t,nh)}{\mathrm{e}^{\alpha t}} v^n + \left(C(t,nh)+\alpha\right)\left(v^n\right)^2 =$$

$$= \left( -\Phi\left(u^n\right) \frac{\left(v^n - v^{n-1}\right)}{h} + \mu \frac{\left(v^{n+1} - v^n\right)}{h} - \frac{dv^n}{dt} \right) v^n \le 0,$$

т.к. $v^n \le 0$, $v^n \le v^{n-1}$, $v^{n+1} \ge v^n$, $\dfrac{dv^n}{dt} \le 0$. Последнее неравенство может быть строгим, только если $t = T$. Итак, мы получили, что

$$v^n \ge -\max_{(t,nh)\in\Pi} \frac{\left|F(t,nh)\right|}{\mathrm{e}^{\alpha t}\left(\alpha + c_0\right)} \ge -\max_{(t,nh)\in\Pi} \frac{\left|F(t,nh)\right|}{\left(\alpha + c_0\right)}.$$

Таким образом, окончательно имеем, что

$$\left|v^n\right| \le \max\left\{ \max_{(t,nh)\in\Pi} \frac{\left|F(t,nh)\right|}{\alpha + c_0},\ \left|v^n\right|_\Gamma \right\}. \qquad (3.7)$$

Так как $w^n = \mathrm{e}^{\alpha t} v^n$ и выполняется (3.7), то

$$\left|w^n\right| \le \max\left\{ \max_{(t,nh)\in\Pi} \frac{\mathrm{e}^{\alpha T}\left|F(t,nh)\right|}{\alpha + c_0},\ \mathrm{e}^{\alpha T}\left|w^n\right|_\Gamma \right\}.$$

Лемма доказана.

**Следствие.** *Пусть*

$$\frac{dw^n}{dt} = -\Phi\left(u^n\right)\frac{\left(w^n - w^{n-1}\right)}{h} + \mu\frac{\left(w^{n+1} - w^n\right)}{h} - C(t,nh)w^n - F(t,nh),$$

*и* $\exists\, c_0 > 0 : \forall\, (t,nh) \in \Pi \to C(t,nh) \ge c_0$. *Тогда*

$$\left|w^n\right| \le \max\left\{\max_{(t,nh)\in\Pi}\frac{\left|F\left(t,nh\right)\right|}{c_0},\ \left|w^n\right|_{\Gamma}\right\}.$$

**Доказательство.** Пусть $w^n$ принимает максимальное значение в некоторой точке $(t,nh) \in \Pi \setminus \Gamma$, не принадлежащей параболической границе, тогда в этой точке

$$C(t,nh)w^n = -\Phi\left(u^n\right)\frac{\left(w^n - w^{n-1}\right)}{h} + \mu\frac{\left(w^{n+1} - w^n\right)}{h} -$$

$$-\frac{dw^n}{dt} - F(t,nh) \le -F(t,nh),$$

т.к. $w^n \ge w^{n-1}$, $w^{n+1} \le w^n$, $\dfrac{dw^n}{dt} \ge 0$. Последнее неравенство может быть строгим, только если $t = T$.

Пусть $w^n$ принимает минимальное значение в некоторой внутренней точке, $(t,nh) \in \Pi \setminus \Gamma$, не принадлежащей параболической границе, тогда в этой точке

$$C(t,nh)w^n = -\Phi\left(u^n\right)\frac{\left(w^n - w^{n-1}\right)}{h} + \mu\frac{\left(w^{n+1} - w^n\right)}{h} -$$

$$-\frac{dw^n}{dt} - F(t,nh) \ge -F(t,nh),$$

т.к. $w^n \le w^{n-1}$, $w^{n+1} \ge w^n$, $\dfrac{dw^n}{dt} \le 0$.

Таким образом, мы получили, что

$$\left|w^n\right| \le \max\left\{ \max_{(t,nh)\in\Pi} \frac{\left|F(t,nh)\right|}{c_0}, \left|w^n\right|_\Gamma \right\}.$$

Следствие доказано.

Будем далее считать, что $\Phi'(u) > 0$, при $u \in [0, 1]$; $\Pi = \{(t,x): 0 \le t \le T, -M \le x \le M\}$. Из работ [8], [78] следует, что при данном $T > 0$, $M > 0$ можно выбрать настолько большим, что максимум $\left|w^n\right|$ на боковой границе будет меньше сколь угодно маленького наперед заданного числа (напомним, что на прямой $t = 0$ $\left|w^n\right| = 0$). Действительно,

$$\forall \sigma > 0 \, \exists M_0 > 0 : \forall x \le -M_0 \to \left|u_0(x) - u_-\right| < \sigma,$$

$$\forall x \ge M_0 \to \left|u_0(x) - u_+\right| < \sigma$$

и в качестве $M$ можно взять любое число не меньшее $\tilde{M}$, где

$$\tilde{M} = M_0 + 2T \max_{u\in[0,1]} \left|\frac{\varphi'(u)}{\eta'(u)}\right| = M_0 + 2T \max_{u=\{0;1\}} \left|\Phi(u) - \mu\right|.$$

Так как $C(t,nh) = \left[ \left( \int_0^1 \Phi'\left(u^n + t\left(u_\varepsilon(nh) - u^n\right)\right) dt \right) \frac{\partial u_\varepsilon\left((n+\theta)h\right)}{\partial x} \right]$, то из [8] следует, что $c_0 \ge 0$ в $\Pi$. Из леммы настоящего параграфа и следствия к ней видно, что самый “плохой” случай – это когда $c_0 = 0$. Этот случай мы и рассмотрим. Беря в (3.6) $\alpha = 1/T$, из (3.4) и утверждения 3 приложения получим, что если $u_0(x)$ имеет ограниченные производные до второго порядка включительно, то

$$\forall n \in Z, t \in [0, T] \to \left|w^n\right| \le T\mathrm{O}(h). \qquad (3.8)$$

Используя результаты работы [79] можно показать, что при $\Phi'(u) > 0$

$$\forall n \in Z, t \in [0, T] \to \left|u_\varepsilon(t,nh) - u(t,nh)\right| \le \sqrt[4]{T}\mathrm{O}\left(\sqrt[4]{h}\right), \qquad (3.9)$$

где $u(t,x)$ – решение задачи Коши (3.10)

$$\frac{\partial u}{\partial t}+\frac{\varphi'(u)}{\eta'(u)}\frac{\partial u}{\partial x}=0,\ u(0,x)=u_0(x). \tag{3.10}$$

При $\Phi'(u)>0$ задача Коши (3.10) имеет в $\pi_\infty$ классическое решение, совпадающее с классическим решением задачи Коши (1.1), (1.2) [стр. 29 – 39, 17], поэтому в формуле (3.9) $u(t,x)$ можно понимать как решение задачи Коши (1.1), (1.2).

Итак, мы доказали, что дифференциально-разностный оператор (3.1) Полтеровича–Хенкина, при определенных условиях, устойчиво аппроксимирует дифференциальный оператор типа Бюргерса (3.2), который, в свою очередь, устойчиво аппроксимирует дифференциальный оператор типа закона сохранения, поэтому [глава 7, § 38, 80] приводимое ниже определение 1 эквивалентно определению обобщенного решения задачи Коши (1.1), (1.2) при условиях:

1) $\Phi(u)>0$, $\mu\geq 0$, $\Phi'(u)>0$, $\eta'(u)=\dfrac{1}{\Phi(u)+\mu}$,

$\varphi'(u)=\dfrac{\Phi(u)-\mu}{\Phi(u)+\mu}$, при $u\in[0,1]$;

2) $u_0(x)$ – является функцией распределения;

3) $u_0(x)$ имеет ограниченные производные до второго порядка включительно.

**Определение 1.** *Ограниченная измеримая функция $u(t,x)$ называется “$\mu$ дифференциально-разностным” решением задачи (1.1), (1.2) в слое $\pi_T$, если $u(t,x)$ есть предел при фиксированном $t>0$ последовательности функций от $x$:*

$u_h(t,x)=\{u^n(t), nh\leq x<(n+1)h, n\in Z\}$ *при* $h\to 0+$, *где* $u^n(t)$ *удовлетворяет (3.1).*

Если $\mu=0$, то при гладкой функции $\Phi(u)>0$ дифференциально-разностное уравнение Полтеровича–Хенкина, описывающие динамику распределения предприятий по уровням эффективности[11] [78], [81], [82]

$$h\frac{du^n}{dt} = -\Phi(u^n)(u^n - u^{n-1}),\ u^n(0) = u_0(nh), \tag{3.11}$$

устойчиво аппроксимирует (1.1) в $L_1$ на конечном отрезке времени и

$$\forall\, T \ge t \ge 0 \to \sup_{n\in Z\setminus K(t,h)} \left|u^n(t) - u(t,nh)\right| = o(h) \text{ при } h \to 0+, \tag{3.12}$$

где $K(t,h)\in Z$ – конечное множество точек, мощность которого может быть ограничена числом, не зависящим от $t \ge 0$ и $h > 0$.

Этот факт можно получить, используя результаты работ [79], [83]. По-видимому, отмеченный результат справедлив и для $T = \infty$, см. статью Хенкина–Шананина// Функ. ан., Т. 50, № 2, (2016).

**Указание.** Обобщая результаты работ Н.С. Бахвалова [84] и Р.В. Разумейко [85], посвященных исследованию устойчивости разностных схем “бегущего счёта” и схем, сводящихся к ним с помощью замены переменных[12], для уравнения (1.1) с $\eta(u) = u$ и $\varphi''(u) > 0$, можно показать, что схема

$$\frac{\eta\left(u_n^{m+1}\right) - \eta\left(u_n^m\right)}{\tau} + \frac{\varphi\left(u_n^m\right) - \varphi\left(u_{n-1}^m\right)}{h} = 0,\ u_n^0 = u_0(0,nh), \text{ при}^{13}$$

$$\frac{\tau}{h}\max_{u\in[0,1]}\left|\frac{\varphi'(u)}{\eta'(u)}\right| \le 1 \text{ (условие Куранта–Фридрихса–Леви [§ 24, 36])}$$

устойчиво аппроксимирует (1.1), (1.2) в слабом смысле на конечном временном отрезке. Из работ Н.Н. Кузнецова [79], [83], [86], [87] следует, что устойчивая аппроксимация (1.1), (1.2) в $L_1$ на конечном временном отрезке имеет место не только схемами “бегу-щего счета", но и схемами "исчезающей вязкости" [79], схемами "сглаживания" [87], дивергентными схемами "распада разрыва" С.К. Годунова[14] [глава 9, 36], [88], [89], причем при более слабых ограничениях на функции $\eta(u)$ и $\varphi(u)$, при произвольной ограниченной измеримой начальной функции.

---

[11] Аналогичное уравнение возникает в микроскопической модели однополосных транспортных потоков G.F. Newell’a, если считать реакцию водителей мгновенной, при этом $u^n$ – скорость $n$-ой машины [стр. 82 – 84, 1].

[12] К таковым относится например схема P.D. Lax’a [12], [18], [34].

[13] Это условие также может быть получено по спектральному признаку, если “заморозить” коэффициенты у уравнения (1.1) [§ 15, § 26, 36].

Можно показать ([79], G. Jennings// Comm. on Pure and Appl. Math., V. 26, (1974); Хенкин–Шананин// Функ. ан., Т. 50, (2016)), что приводимое ниже определение 2 эквивалентно определению обобщенного решения задачи Коши (1.1), (1.2) при условиях:

1) $\Phi(u)>0$, $\mu=0$, $\eta'(u)=\dfrac{1}{\Phi(u)}$, $\varphi'(u)\equiv 1$, при $u\in[0,1]$;

2) $u_0(x)$ – является непрерывной функцией распределения с конечным носителем.

3) в точках перехода с одной волны на другую $\Phi'(u_k)>0$.

**Определение 2 (А.А. Шананин, Г.М. Хенкин, В.М. Полтерович).** *Ограниченная измеримая функция $u(t,x)$ называется дифференциально-разностным решением задачи (1.1), (1.2) в слое $\pi_\infty$, если $u(t,x)$ удовлетворяет (3.12) и*

$$\forall\, 0\le t\le\infty\to\sum_{n=-\infty}^{\infty}\left|u^n(t)-u(t,nh)\right|\to 0 \text{ при } h\to 0+.$$

Резюмируя результаты этого параграфа, приходим к следующему утверждению, являющемуся обобщением утверждения 2 § 2.

**Утверждение.** *Определениям 1 – 8 § 1; 6, 7 § 2 и 1, 2 § 3 удовлетворяет одна и та же единственная (с точностью до почти всюду по $x$ при фиксированном $t>0$) функция $u(t,x)$, которую будем называть обобщенным решением. Кроме того, имеет место непрерывная зависимость $u(t,x)$ от $u_0(x)$:*

$$\text{если } u_0^k(x)\xrightarrow{L_1}u_0(x),\ \text{то } u^k(t,x)\xrightarrow{L_1}u(t,x).$$

[14] Заметим, что в случае $\Phi(u)>0$, $u_0(x)$ – функция распределения, схема "распада разрыва" С.К. Годунова совпадает со схемой бегущего счёта.

Более подробно о разностных схемах для уравнения (1.1) написано в [глава 3, 9], [глава 2, 90]. См. также Holden H., Risebro N.H. Front tracking for hyperbolic conservation laws. Springer, 2007.

Результаты этого параграфа допускают ряд обобщений (см. сноску 7, работы Serre D. $L^1$ stability of shock waves in scalar conservation laws, in: Evolutionary Equations// Handbook of Differential Equations, North-Holland, Amsterdam, v.1, (2004), pp.473-553; Гасников А.В. О гипотезе Полтеровича–Хенкина и её обобщении// Сборник трудов II Всеросс. научн. конф. "ЭКОМОД-2007", посвященной 90-летию со дня рождения Н.Н. Моисеева, Киров, 9-15 июля 2007. Киров: Изд-во ВятГУ, (2007), стр.77-85; и цитированную в них литературу).

## ПРИЛОЖЕНИЕ: ОЦЕНКИ ПРОИЗВОДНЫХ РЕШЕНИЯ КВАЗИЛИНЕЙНОГО УРАВНЕНИЯ ПАРАБОЛИЧЕСКОГО ТИПА

Рассмотрим следующую начальную задачу Коши для квазилинейного уравнения параболического типа [3]

$$\frac{\partial \eta(u_\varepsilon)}{\partial t} + \frac{\partial \varphi(u_\varepsilon)}{\partial x} = \varepsilon \frac{\partial^2 u_\varepsilon}{\partial x^2}\ ,\ \varepsilon > 0, \quad (1)$$

$$u_\varepsilon\big|_{t=0} = u_0(x), \quad (2)$$

где $u_0(x)$ – произвольная ограниченная ($u_0(x) \in [a, b]$) измеримая функция. Равенство (2) понимается в слабом смысле. Будем предполагать, что функции $\eta(u)$, $\varphi(u)$, определенные на $U = [a, b]$, дважды непрерывно дифференцируемы, а также

$$\exists\ \bar{a} > 0:\ \eta'(u) \ge \bar{a} \text{ на } U = [a, b]. \quad (3)$$

Из результатов работ [4], [5], [12] имеем:

**Теорема (О.А. Олейник).** *Решение задачи Коши (1), (2) при фиксированном $\varepsilon > 0$ существует, единственно и ограничен-*

*но. Более того, в любой полосе* $0 < t_0 < t < T$ *оно имеет ограниченные производные, входящие в уравнение (1).*

Из [13] следует:

**Утверждение 1.** *Для всех* $x$ *и* $t \geq t_0 > 0$

$$\left| \varepsilon \frac{\partial u_\varepsilon}{\partial x} \right| \leq C_1,$$

*где* $C_1$ *не зависит от* $\varepsilon$.

Из [5], [7], [8], [11] следует:

**Лемма (Обобщенный принцип максимума).** *Пусть* $u(t,x)$ – *непрерывная при* $t \geq 0$ *функция, имеющая при* $t > 0$ *непрерывные производные* $\frac{\partial u(t,x)}{\partial x}$, $\frac{\partial^2 u(t,x)}{\partial x^2}$, $\frac{\partial u(t,x)}{\partial t}$ *и такая, что при* $u(t,x) \leq 0$ *выполняется неравенство*

$$L_\varepsilon u(t,x) = \varepsilon(t,x)\frac{\partial^2 u}{\partial x^2} - r(t,x)\frac{\partial u}{\partial t} + a(t,x)\frac{\partial u}{\partial x} + c(t,x)u \leq 0,$$

*где* $c(t,x) \leq 0$; $r(t,x) \geq \tilde{r} > 0$; $\varepsilon(t,x) \geq 0$ – *ограниченная функция при всех* $0 \leq t \leq T$ *для любого фиксированного* $T$; $a(t,x)$ – *функция, ограниченная сверху при* $x > 0$, *ограниченная снизу при* $x < 0$ *и при всех* $0 \leq t \leq T$ *для любого фиксированного* $T$. *Тогда, если*

$$u(t,x) \geq -A(t)\sqrt{x^2+1}, \tag{4}$$

*где* $A(t)$ – *непрерывная функция, и* $u(0,x) \geq 0$, *то при* $t \geq 0$

$$u(t,x) \geq 0.$$

**Замечание 1.** Если функция $u_\varepsilon(t,x)$, для которой справедливо (4) (это, в частности, справедливо для ограниченной $u(t,x)$), удовлетворяет при $t > 0$ (1) и $u_\varepsilon(0,x) \in [a,b]$, то

$$\forall\, t \geq 0,\, x \in R^1 \rightarrow u_\varepsilon(t,x) \in [a,b].$$

Если предположить, что функции $\eta(u)$, $\varphi(u)$, определенные на $U = [a,b]$, $k+1$ раз непрерывно дифференцируемы, то с помо-щью оценок аналогичных оценкам С.Н. Бернштейна [5], [91] можно показать, что уравнение (1) можно дифференцировать $k$ раз.

1) $L_\varepsilon^1 v_1 = \varepsilon_1(u_\varepsilon)\dfrac{\partial^2 v_1}{\partial x^2} - \dfrac{\partial v_1}{\partial t} + a_1(u_\varepsilon)\dfrac{\partial v_1}{\partial x} + b_1(u_\varepsilon)\left(v_1\right)^2 = 0$,

$$a_1(u) = -\left(\frac{\varphi'(u)}{\eta'(u)} + \varepsilon\frac{\eta''(u)}{(\eta'(u))^2}\frac{\partial u_\varepsilon}{\partial x}\right) = -a(u),\ b_1(u) = \mp\left(\frac{\varphi'(u)}{\eta'(u)}\right)' = \mp b(u),$$

$$\varepsilon_1(u) = \frac{\varepsilon}{\eta'(u)} > 0,\ v_1 = \pm\frac{\partial u_\varepsilon}{\partial x}.$$

2) $L_\varepsilon^2 v_2 + b_2(u_\varepsilon)\left(v_2\right)^2 = \varepsilon_2(u_\varepsilon)\dfrac{\partial^2 v_2}{\partial x^2} - \dfrac{\partial v_2}{\partial t} + a_2(u_\varepsilon)\dfrac{\partial v_2}{\partial x} + c_2(u_\varepsilon)v_2 +$

$$+b_2(u_\varepsilon)\left(v_2\right)^2 = F_2(t,x),\ a_2(u) = -\left(\frac{\varphi'(u)}{\eta'(u)} + 2\varepsilon\frac{\eta''(u)}{(\eta'(u))^2}\frac{\partial u_\varepsilon}{\partial x}\right),$$

$$c_2(u) = -3b_1(u)\frac{\partial u_\varepsilon}{\partial x} - \varepsilon\frac{\eta'''(u)\eta'(u) - 2\left(\eta''(u)\right)^2}{(\eta'(u))^3}\left(v_1\right)^2,$$

$$b_2(u) = \mp\varepsilon\frac{\eta''(u)}{(\eta'(u))^2},\ \varepsilon_2(u) = \varepsilon_1(u) = \frac{\varepsilon}{\eta'(u)} > 0,\ v_2 = \pm\frac{\partial^2 u_\varepsilon}{\partial x^2},$$

$$F_2(t,x) = \pm b_1'(u)\left(\frac{\partial u_\varepsilon}{\partial x}\right)^3 = P_2(v_1).$$

3) $L_\varepsilon^3 v_3 = \varepsilon_3(u_\varepsilon)\dfrac{\partial^2 v_3}{\partial x^2} - \dfrac{\partial v_3}{\partial t} + a_3(u_\varepsilon)\dfrac{\partial v_3}{\partial x} + c_3(u_\varepsilon)v_3 = F_3(t,x)$,

$$a_3(u) = -\left(\frac{\varphi'(u)}{\eta'(u)} + 3\varepsilon\frac{\eta''(u)}{(\eta'(u))^2}\frac{\partial u_\varepsilon}{\partial x}\right),\ c_3(u) = -4b_1(u)\frac{\partial u_\varepsilon}{\partial x} -$$

$$-3\varepsilon\frac{\eta'''(u)\eta'(u) - 2\left(\eta''(u)\right)^2}{(\eta'(u))^3}\left(v_1\right)^2 - 4\varepsilon\frac{\eta''(u)}{(\eta'(u))^2}\frac{\partial^2 u_\varepsilon}{\partial x^2},\ \varepsilon_3(u) = \frac{\varepsilon}{\eta'(u)},$$

$$v_3 = \pm\frac{\partial^3 u_\varepsilon}{\partial x^3},\ F_3(t,x) = \pm b_1''(u)\left(\frac{\partial u_\varepsilon}{\partial x}\right)^4 \pm 6b_1'(u)\left(\frac{\partial u_\varepsilon}{\partial x}\right)^2\frac{\partial^2 u_\varepsilon}{\partial x^2} \pm$$

$$\pm 3b_1(u)\left(\frac{\partial^2 u_\varepsilon}{\partial x^2}\right)^2 + \varepsilon M\left(\eta'',\eta'''\right) = P_3(v_1,v_2)$$, где $M\left(0,0\right)=0$, т.е. если

$\eta(u)$ – линейная функция, то последнего слагаемого не будет.

………………………………………………………………………

k) $L_\varepsilon^k v_k = \varepsilon_k(u_\varepsilon)\frac{\partial^2 v_k}{\partial x^2} - \frac{\partial v_k}{\partial t} + a_k(u_\varepsilon)\frac{\partial v_k}{\partial x} + c_k(u_\varepsilon)v_k = F_k(t,x)$,

$$a_k(u) = -\left(\frac{\varphi'(u)}{\eta'(u)} + k\varepsilon\frac{\eta''(u)}{(\eta'(u))^2}\frac{\partial u_\varepsilon}{\partial x}\right),\ c_k(u) = -\left(k+1\right)b_1(u)\frac{\partial u_\varepsilon}{\partial x} -$$

$$-\varepsilon\frac{\left(k-1\right)k}{2}\frac{\eta'''(u)\eta'(u) - 2\left(\eta''(u)\right)^2}{(\eta'(u))^3}\left(v_1\right)^2 - \varepsilon\left(\frac{\left(k-1\right)k}{2}+1\right)\frac{\eta''(u)}{(\eta'(u))^2}\frac{\partial^2 u_\varepsilon}{\partial x^2},$$

$$\varepsilon_k(u) = \frac{\varepsilon}{\eta'(u)} > 0,\ v_k = \pm\frac{\partial^k u_\varepsilon}{\partial x^k},$$

$$F_k(t,x) = P_k(v_1,\dots,v_{k-1}) = \frac{\partial}{\partial x}P_{k-1}(v_1,\dots,v_{k-2}) - v_{k-1}\frac{\partial}{\partial x}c_k(u_\varepsilon).$$

С помощью обобщенного принципа максимума из 1) – k) можно получать оценки на производные по $x$ решения (1).

**Утверждение 2.** *Пусть $\varphi\left(\eta^{-1}\left(y\right)\right)$ – строго выпуклая по $y$ функция, т.е. $\left(\varphi'(u)/\eta'(u)\right)' > 0$, тогда для всех $x$ и $t \ge t_0 > 0$*

$$\frac{\partial u_\varepsilon}{\partial x} \le \frac{E_1}{t},$$

*где $E_1$ не зависит от $\varepsilon$.*

**Доказательство.** Разделим (1) на $\eta'(u) \ge \bar{a} > 0$, и то, что получится, продифференцируем по $x$. Как указывалось выше, мы вправе это сделать. Производная $-\frac{\partial u_\varepsilon(t,x)}{\partial x}$, которую мы будем обозначать $\tilde{v}(t,x)$, удовлетворяет уравнению

$$\frac{\partial\tilde{v}}{\partial t} + a(u_\varepsilon)\frac{\partial\tilde{v}}{\partial x} - b(u_\varepsilon)\tilde{v}^2 - \frac{\varepsilon}{\eta'(u_\varepsilon)}\frac{\partial^2\tilde{v}}{\partial x^2} = 0,$$

где введены обозначения $a(u)=\left(\frac{\varphi'(u)}{\eta'(u)}+\frac{\eta''(u)}{(\eta'(u))^2}\left(\varepsilon\frac{\partial u_\varepsilon}{\partial x}\right)\right)$ и $b(u)=\left(\frac{\varphi'(u)}{\eta'(u)}\right)'$. В силу условия (3) и достаточной гладкости функций $\varphi(u)$ и $\eta(u)$ имеем, что $a(u)$ и $b(u)$ - ограниченные функции и функция $b(u)$ не зависит от $\varepsilon$, поэтому

$$\exists\,\mu>0:\forall\,\varepsilon>0,\ u\in U\to b(u)\geq\mu.$$

Более того, из утверждения 1 имеем, что $a(u)$ равномерно ограничена по $\varepsilon$. Можно показать, подобно тому, как это было сделано в работе [12], что

$$\forall\,t_0>0\,\exists\,E(t_0)\geq\frac{1}{\mu}:\forall\,E_1\geq E(t_0),\,\varepsilon>0,\,x\in R^1\to\frac{E_1}{t_0}+\tilde v(t_0,x)\geq 0.$$

Далее, когда $\frac{E}{t}+\tilde v(t,x)\leq 0$, имеем

$$L^1_\varepsilon\left(\tilde v+\frac{E}{t}\right)\equiv\frac{\varepsilon}{\eta'(u_\varepsilon)}\frac{\partial^2\left(\tilde v+\frac{E_1}{t}\right)}{\partial x^2}-\frac{\partial\left(\tilde v+\frac{E_1}{t}\right)}{\partial t}-a(u)\frac{\partial\left(\tilde v+\frac{E_1}{t}\right)}{\partial x}+$$

$$+b(u)\left(\tilde v+\frac{E_1}{t}\right)^2\leq\frac{E_1}{t^2}+b(u)\left(2\frac{E_1}{t}\tilde v+\frac{E_1^2}{t^2}\right)\leq\frac{E_1}{t^2}-b(u)\frac{E_1^2}{t^2}\leq 0,$$

так как $L^1_\varepsilon\tilde v=0$.

$$L_\varepsilon\left(\tilde v+\frac{E_1}{t}\right)\equiv\frac{\varepsilon}{\eta'(u_\varepsilon)}\frac{\partial^2\left(\tilde v+\frac{E_1}{t}\right)}{\partial x^2}-\frac{\partial\left(\tilde v+\frac{E_1}{t}\right)}{\partial t}-a(u)\frac{\partial\left(\tilde v+\frac{E_1}{t}\right)}{\partial x}\leq$$

$$\leq -b(u)\left(\tilde v+\frac{E_1}{t}\right)^2\leq 0.$$

По теореме О.А. Олейник $\tilde v$ – ограниченная функция в $S_{T,\infty}\setminus\mathring{S}_{t_0,\infty}$ для любого фиксированного $t_0>0$, пускай и неравномерно по $\varepsilon$ и по $t_0>0$. Поэтому, при фиксированных $\varepsilon$ и $t_0>0$, рассматривая нашу задачу Коши как задачу Коши с начальным условием не при $t=0$, а при $t=t_0$, мы будем находиться в условиях обобщенного принципа максимума, и, следовательно,

$$\tilde{v}+\frac{E_1}{t}\geq 0 \text{ – для всех } x \text{ и } t\geq t_0>0.$$

Но это означает, что

$$\frac{\partial u_\varepsilon}{\partial x}\leq\frac{E_1}{t} \text{ – для всех } x \text{ и } t\geq t_0>0,$$

причем $E$ может зависеть только от $t_0>0$. Утверждение 2 доказано.

**Следствие.** *Пусть* $\left(\varphi'(u)/\eta'(u)\right)'>0$ *и* $u_0'(x)\leq K$, *тогда для всех* $x$ *и* $t\geq 0$

$$\frac{\partial u_\varepsilon}{\partial x}\leq K \quad .$$

Рассмотрим случай, когда $\eta(u)=u$. Тогда

$$L_\varepsilon^k v_k=\varepsilon\frac{\partial^2 v_k}{\partial x^2}-\frac{\partial v_k}{\partial t}+a_k(u_\varepsilon)\frac{\partial v_k}{\partial x}+c_k(u_\varepsilon)v_k=F_k(t,x),\ a_k(u)=-\varphi'(u),$$

$$c_k(u)=-\left(k+1\right)\varphi''(u)\frac{\partial u_\varepsilon}{\partial x},\ v_k=\pm\frac{\partial^k u_\varepsilon}{\partial x^k},$$

$$F_k(t,x)=P_k(v_1,...,v_{k-1})=\frac{\partial}{\partial x}F_{k-1}(t,x)-v_{k-1}\frac{\partial}{\partial x}c_k(u_\varepsilon),$$

$$F_2(t,x)=\pm\varphi'''(u)\left(\frac{\partial u_\varepsilon}{\partial x}\right)^3,\ k\geq 2.$$

**Утверждение 3.** *Пусть* $\varphi\in C^3$, $\varphi''(u)>0$, $\eta(u)=u$ *при* $u\in\left[a,b\right]$ *и* $u_0'(x)\geq 0$, *тогда*

$$\left|\frac{\partial^2 u_\varepsilon}{\partial x^2}\right|\leq\max\left\{\frac{E_2}{t_0^2},\max_x\left|\frac{\partial^2 u_\varepsilon}{\partial x^2}\left(t_0,x\right)\right|\right\},$$

*где* $E_2 = \dfrac{(E_1)^2}{3} \max\limits_{u\in[a,b]} \dfrac{\varphi'''(u)}{\varphi''(u)}$ *– не зависит от* $\varepsilon > 0$ *и* $t_0 > 0$*, а* $E_1$ *из утверждения 2.*

**Доказательство.** Имеем, $k = 2$. Положим $\tilde{v}_2 = \pm\dfrac{\partial^2 u_\varepsilon}{\partial x^2}$.

$$L_\varepsilon^2(\tilde{v}_2 + D_2) = \varepsilon\frac{\partial^2 \tilde{v}_2}{\partial x^2} - \frac{\partial \tilde{v}_2}{\partial t} + a_2(u_\varepsilon)\frac{\partial \tilde{v}_2}{\partial x} + c_2(u_\varepsilon)(\tilde{v}_2 + D_2) =$$

$$= F_2(t,x) + c_2(u_\varepsilon)D_2 = -3\varphi''(u)\frac{\partial u_\varepsilon}{\partial x}D_2 \pm \varphi'''(u)\left(\frac{\partial u_\varepsilon}{\partial x}\right)^3.$$

По условию $\varphi''(u) > 0$ при $u \in [a,b]$ и $u_0'(x) \ge 0$, тогда из утверждения 2 и следствия к нему имеем, что $0 \le \dfrac{\partial u_\varepsilon}{\partial x} \le \dfrac{E_1}{t}$, $t \ge t_0 > 0$ и существует такое $D_2 = \max\left\{\dfrac{E_2}{t_0^2}, \max\limits_{x\in R^1}\left|\dfrac{\partial^2 u_\varepsilon}{\partial x^2}(t_0, x)\right|\right\}$, где $E_2 = \dfrac{(E_1)^2}{3}\max\limits_{u\in[a,b]}\dfrac{\varphi'''(u)}{\varphi''(u)}$ – не зависит от $\varepsilon > 0$ и $t_0 > 0$, для которого $L_\varepsilon^2(\tilde{v}_2 + D_2) = -3\varphi''(u)\dfrac{\partial u_\varepsilon}{\partial x}D_2 \pm \varphi'''(u)\left(\dfrac{\partial u_\varepsilon}{\partial x}\right)^3 \le 0$. Из обобщенного принципа максимума отсюда следует, что $\left|\dfrac{\partial^2 u_\varepsilon}{\partial x^2}\right| \le D_2$. Утверждение 3 доказано.

**Замечание 2.** Если в условиях утверждения 3 $\eta'(u) > 0$, то удается только получить равномерную по времени и $\varepsilon$ оценку для второй производной решения (1), (2) сверху, когда $\eta''(u) \ge 0$ и снизу, когда $\eta''(u) \le 0$, начиная с некоторого момента времени. Доказательство схоже с доказательством утверждений 2 и 3.

Подробнее ознакомиться с использованной техникой можно, например, по монографии [92].

## ЛИТЕРАТУРА

1. *Уизем Дж.* Линейные и нелинейные волны// М.: Мир, (1977).
2. *Гельфанд И.М.* Некоторые задачи теории квазилинейных уравнений// УМН, т.14, вып. 2(86), (1959), стр.87-158.
3. *Serre D.* System of conservation laws: A challenge for the XXIst century, in: B. Enquist, W. Schmid (Eds.), Mathematics Unlimited – 2001 and Beyond// Springer-Verlag, Berlin, New York, (2001), pp.1061-1080.
4. *Олейник О.А.* Краевые задачи для уравнений с частными производными с малым параметром при старших производных и задача Коши для нелинейных уравнений в целом// Докт. дисс., М.: Мехмат, (1954), стр.45-182.
5. *Олейник О.А., Вентцель Т.Д.* Первая краевая задача и задача Коши для квазилинейных уравнений параболического типа// Матем. сб., т.41(83), № 1, (1957), стр.105-128.
6. *Ballou D.P.* Solution to nonlinear hyperbolic Cauchy problems without convexity condition// Trans. Amer. Math. Soc., v.152, № 2, (1970), pp.441-460.
7. *Гасников А.В.* Асимптотическое по времени поведение решения квазилинейного уравнения параболического типа// ЖВМ и МФ, т.46, № 12, (2006), стр.2237-2255.
8. *Weinberger H.F.* Long-time behavior for regularized scalar conservation law in absence of genuine nonlinearity// Ann. Inst. H. Poincaré Anal. Non Linéaire, (1990), pp.407-425.
9. *Рождественский Б.Л., Яненко Н.Н.* Системы квазилинейных уравнений и их приложения к газовой динамике// М.: Наука, (1978).
10. *Кружков С.Н.* Квазилинейные уравнения первого порядка со многими независимыми переменными// Матем. сб., т.81(123), № 2, стр.228-255; Труды С.Н. Кружкова, М.: ФИЗМАТЛИТ, (2000), стр.61-98.
11. *Ильин А.М., О.А. Олейник О.А.* Асимптотическое поведение решений задачи Коши для некоторых квазилинейных уравне-

ний при больших значениях времени// Матем. сб., т.51(93), № 2, (1960), стр.191-216.

12. *Олейник О.А.* Разрывные решения нелинейных дифференциальных уравнений// УМН, т.12, вып. 3(75), (1957), стр.3-73.
13. *Олейник О.А.* О построении обобщенного решения задачи Коши для квазилинейного уравнения первого порядка путем введения “исчезающей вязкости”// УМН, т.14, вып. 2(86), (1959), стр.159-164.
14. *Ландау Л.Д., Лифшиц Е.М.* Гидродинамика// М.: Наука, (1988).
15. *Liu T.-P.* On shock wave theory// Taiwanese journal of math., v.4, № 1, (2000), pp.9-20.
16. *Ладыженская О.А.* Шестая проблема тысячелетия: уравнение Навье-Стокса, существование и гладкость// УМН, т.58, вып. 2(350), (2003), стр.45-78.
17. *Горицкий А.Ю., Кружков С.Н., Чечкин Г.А.* Уравнения с частными производными первого порядка (учебное пособие)// М.: Мехмат, (1999).
18. *Lax P.D.* Weak solution of nonlinear hyperbolic equation and their numerical computation// Comm. Pure Appl. Math., v.7, № 1, (1954), pp.159-193.
19. *Петровский И.Г.* Лекции об уравнениях в частных производных// М.: Гостехиздат, (1953).
20. *Hopf E.* The partitial differential equation $u_t + uu_x = \mu u_{xx}$// Comm. Pure Appl. Math, v.3, № 3, (1950), pp.201-230.
21. *Соболев С.Л.* Некоторые применения функционального анализа в математической физике// М.: Наука, (1988).
22. *Курант Р., Фридрихс К.* Сверхзвуковое течение и ударные волны// М.: ИЛ, (1950).
23. *Олейник О.А.* О задаче Коши для нелинейных уравнений в классе разрывных функций// ДАН СССР, т.95, № 3 (1954), стр.451-455.
24. *Натансон И.П.* Теория функций вещественной переменной// М.: Наука, (1974).

25. *Канторович Л.В., Акилов Г.П.* Функциональный анализ// Издательство bhv®, Санкт-Петербург, (2004).
26. *Олейник О.А.* Задача Коши для нелинейных дифференциальных уравнений первого порядка с разрывными начальными условиями// Труды Московского математического общества, т.5, (1956), стр.433-454.
27. *Олейник О.А.* О единственности и устойчивости обобщенного решения задачи Коши для квазилинейного уравнения.// УМН, т.14, вып. 2(86), (1959), стр.165-170.
28. *Олейник О.А.* Об одном классе разрывных решений квазилинейных уравнений первого порядка// Научные доклада высшей школы, Физико-математические науки, № 3, (1958), стр.91-98.
29. *Liu T.-P.* Invariants and asymptotic behavior of solution of a conservation law// Proc. Amer. Math. Soc., v.71, № 2, (1978), pp.227-231.
30. *Liu T.-P.* Admissible solutions of hyperbolic conservation laws// Mem. Amer. Math. Soc., v.30, (1981), pp.1-78.
31. *Cheng K.-S.* Asymptotic behavior of solution of a conservation law without convexity condition// J. Diff. Equat., v.40, № 3, (1981), pp.343-376.
32. *Калашников А.С.* Построение обобщенных решений квазилинейных уравнений первого порядка без условия выпуклости как предел решений параболических уравнений с маленьким параметром// ДАН СССР, т.127, № 1, (1959), стр.27-30.
33. *Рождественский Б.Л.* Новый метод решения задачи Коши в целом для квазилинейных уравнений// ДАН СССР, т.138, № 2, (1961), стр.309-312.
34. *Lax P.D.* Hyperbolic system of conservation laws// Comm. Pure Appl. Math., v.10, № 4, (1957), pp.537-566.
35. *Зельдович Я.Б., Райзер Ю.П.* Физика ударных волн и высокотемпературных гидродинамических явлений// М.: Наука, (1966).

36. *Годунов С.К., Рябенький В.С.* Разностные схемы// М.: Наука, (1977).
37. *Годунов С.К.* Проблема обобщенного решения в теории квазилинейных уравнений в газовой динамике// УМН, т.17, вып. 3(105), (1962), стр.147-158.
38. *Куликовский А.Г., Свешникова Е.И.* Нелинейные волны в упругих средах// М.: Моск. лицей, (1998).
39. *Кружков С.Н.* Обобщенные решения задачи Коши в целом для нелинейных уравнений первого порядка в классе разрывных функций// ДАН СССР, т.225, № 5, (1969), стр.25-28.
40. *Панов Е.Ю.* О единственности решения задачи Коши для квазилинейного уравнения первого порядка с одной допустимой строго выпуклой энтропией// Матем. заметки, т.55, № 5, (1994), стр.116-129; *Панов Е.Ю.* О мерозначных решениях задачи Коши для квазилинейного уравнения первого порядка// Известия РАН, серия математическая, т.60, № 2, (1996), стр.107-148.
41. *Perthame B., Tadmor E.* A kinetic equation with kinetic entropy functions for scalar conservation laws// Commun. Math. Phys., v.136, № 3, (1990), pp.501-517.
42. *Henkin G.M., Shananin A.A.* Asymptotic behavior of solutions of the Cauchy problem for Burgers type equations// Journal de mathématiques purés et appliquées, v.83, (2004), pp.1457-1500.
43. *Henkin G.M., Shananin A.A., Tumanov A.E.* Estimates for solution of Burgers type equations and some applications// Journal de mathématiques purés et appliquées, v.84, (2005), pp.717-752.
44. *Кузнецов Н.Н.* О некоторых асимптотических свойствах обобщенного решения задачи Коши для квазилинейного уравнения первого порядка// УМН, т.14, вып. 2(86), (1959), стр.203-209.
45. *Кружков С.Н., Петросян Н.С.* Асимптотическое поведение решений задачи Коши для нелинейных уравнений первого порядка// УМН, т.42, № 5(257), (1987), стр.3-40.
46. *Петросян Н.С.* Об асимптотическом поведении решения задачи Коши для квазилинейного уравнения первого порядка

при $t \to +\infty$// Динамика сплошной среды, Новосибирск, вып. 36, (1978), стр.86-96.

47. *Петросян Н.С.* Об асимптотических свойствах решения задачи Коши для квазилинейного уравнения первого порядка// Дифференциальные уравнения, т.20, № 3, (1984), стр.502-508.
48. *Dafermors C.M.* Characteristics in hyperbolic conservation laws. A study of structure and the asymptotic behavior of solution// Heriot-Watt symposium, v.1, Research notes in math., London: Pitman, v.17, (1977), pp.1-58.
49. *Эванс Л.К.* Уравнения с частными производными// Университетская серия, том 7, Новосибирск, (2003).
50. *DiPerna R.J.* Decay and asymptotic behavior of solution to nonlinear hyperbolic system of conservation laws// Ind. Univ. math. journ., v.24, № 11, (1975), pp.1047-1071.
51. *Петровский И.Г.* Лекции по теории обыкновенных дифференциальных уравнений// М.: Наука, (1964).
52. *Субботин А.И.* Обобщенные решения уравнений в частных производных. Перспективы динамической оптимизации// Институт Компьютерных Исследований, Москва, Ижевск, (2003).
53. *Кружков С.Н.* Обобщенные решения нелинейных уравнений первого порядка со многими независимыми переменными I// Матем. сб., т.70(112), (1966), стр.394-415.
54. *Кружков С.Н.* Обобщенные решения нелинейных уравнений первого порядка со многими независимыми переменными II// Матем. сб., т.72(114), (1967), стр.108-134.
55. *Кружков С.Н.* О минимаксном представлении решений нелинейных уравнений первого порядка// Функ. ан. и его приложения, вып. 2, № 3, (1969), стр.57-66; Труды С.Н. Кружкова, М.: ФИЗМАТЛИТ, (2000), стр.46-60.
56. *Crandall M.G., Lions P.L.* Viscosity solution of Hamilton-Jacobi equation// Trans. Amer. Math. Soc., v.277, № 1, (1983), pp.1-42.
57. *Маслов В.П., Самборский С.Н.* Существование и единственность решений стационарных уравнений Гамильтона-Якоби и

Беллмана. Новый подход// ДАН, т.324, № 6, (1992), стр.1143-1148.
58. *Маслов В.П., Колокольцев В.Н.* Идемпотентный анализ и его приложения в оптимальном управлении// М.: Наука, (1994).
59. *Maslov V.P., Kolokoltsev V.N.* Idempotent analysis and its application// London: Kluwer Acad. Pull., (1997).
60. *Krasovskiĭ N.N., Subbotin A.I.* Game-Theoretical control problems// Springer-Verlag, New York, 1988.
61. *Меликян А.А.* Сингулярные характеристики уравнений в частных производных первого порядка// ДАН, т.82, № 2, (1996), стр. 203-217.
62. *Меликян А.А.* Уравнения распространения слабого разрыва решения вариационной задачи// Труды института математики и механики УрО РАН, т.6, № 2, (2000), стр.446-459.
63. *Melikyan A.A.* Generalized characteristic of first order PDEs: Application in optimal control and differential games// Boston, Birkhäuser, (1998).
64. *Evans L.C., Sougandis P.E.* Differential games and representation formulas for solutions of Hamilton-Jacobi-Isaacs equations// Indiana Univ. Math. Journ., v.33, (1984), pp.773-797.
65. *Hopf E.* General solution of nonlinear equation of first order// Journal Math. Mech., v.14, (1965), pp.951-973.
66. *Bardi M., Evans L.C.* On Hopf's formulas for solution of Hamilton-Jacobi equations// Nonlinear Anal., Theory, Meth. Appl., v.8, № 11, (1984), pp.1373-1381.
67. *Ибрагимов Н.Х.* Группы преобразований в математической физике// М.: Наука, (1983).
68. *Кузнецов Н.Н., Рождественский Б.Л.* Построение обобщенного решения задачи Коши для квазилинейного уравнения// УМН, т.14, вып. 2(86), (1959), стр.211-215.
69. *Кузнецов Н.Н., Рождественский Б.Л.* К вопросу о построение обобщенного решения задачи Коши для квазилинейного уравнения// УМН, т.20, вып. 1(121), (1965), стр.209-212.
70. *Субботин А.И.* Минимаксные неравенства и уравнение Гамильтона-Якоби// М.: Наука, (1991).

71. *Галлеев Э.М., Тихомиров В.М.* Оптимизация: теория, примеры, задачи// УРСС, Москва, (2000).
72. *Куржанский А.Б.* Управление и наблюдение в условиях неопределенности// М.: Наука, (1977).
73. *Магарил-Ильяев Г.Г., Тихомиров В.М.* Выпуклый анализ и его приложения// УРСС, Москва, (2003).
74. *Демьянов В.Ф.* Минимакс, дифференцируемость по направлениям// Издательство ЛГУ, Ленинград, (1974).
75. *Рокафеллар Р.* Выпуклый анализ// М.: Мир, (1973).
76. *Рублев И.В.* Обобщение формулы Хопфа для неавтоноиного уравнения Гамильтона-Якоби// МГУ ВМиК, Прикладная математика и информатика № 3, Москва, Диалог МГУ, (1999), стр.81-89.
77. *Гельман Л.М., Левин М.И., Полтерович В.М., Спивак В.А.* Отраслевые проблемы Моделирование динамики распределения предприятий отрасли по уровням эффективности (на примере черной металлургии)// Экономика и математические методы, т.29, вып. 3, (1993), стр.460-469.
78. *Henkin G.M., Polterovich V.M.* A difference-differential analogue of the Burgers equation and some models of economic development// Discrete and continuous dynamic systems, v.5, № 4, (1999), pp.697-728.
79. *Кузнецов Н.Н.* Точность некоторых приближенных методов расчета слабых решений квазилинейного уравнения первого порядка// ЖВМ и МФ, т.16, № 6, (1976), стр.1489-1502.
80. *Треногин В.А.* Функциональный анализ// М.: ФИЗМАТЛИТ, (2002).
81. *Полтерович В.М., Хенкин Г.М.* Математический анализ экономических моделей Эволюционная модель взаимодействия процессов создания и заимствования технологий// Экономика и математические методы, т.24, вып. 6, (1988), стр.1071-1083.
82. *Henkin G.M., Polterovich V.M.* Shumpetrian dynamics as nonlinear wave theory// Journal of Mathematical Economics, v.20, (1991), pp.551-590.

83. *Кузнецов Н.Н.* Об устойчивых методах решения квазилинейного уравнения первого порядка в классе разрывных функций// ДАН СССР, т.225, № 5, (1975), стр.1009-1012.
84. *Бахвалов Н.С.* Оценка погрешности численного интегрирования квазилинейного уравнения первого порядка// ЖВМ и МФ, т.1, № 5, (1961), стр.771-783.
85. *Разумейко Р.В.* Оценка погрешности численного интегрирования квазилинейного уравнения первого порядка// Матем. зам., т.13, № 2, (1973), стр.207-215.
86. *Кузнецов Н.Н.* Об одном конечно – разностном методе решения задачи Коши для квазилинейного уравнения первого порядка// ЖВМ и МФ, т.17, № 3, (1977), стр.676-689.
87. *Кузнецов Н.Н.* О применении метода сглаживания к некоторым системам гиперболических квазилинейных уравнений// ЖВМ и МФ, т.13, № 1, (1973), стр.92-102.
88. *Годунов С.К.* Разностный метод численного расчета разрывных решений уравнений гидродинамики// Матем. сб., т.47(89), № 3, (1959), стр.271-306.
89. *Годунов С.К.* Оценка невязок для приближенных решений простейших уравнений газовой динамики// ЖВМ и МФ, т.1, № 4, (1961), стр.623-637.
90. *Куликовский А.Г., Погорелов Н.В., Семенов А.Ю.* Математические вопросы численного решения гиперболических систем уравнений// М.: ФИЗМАТЛИТ, (2001).
91. *Бернштейн С.Н.* Ограничение модулей последовательных производных решений уравнений параболического типа// ДАН СССР, т.18, № 7, (1938), стр.385-388.
92. *Ладыженская О.А., Солонников В.А., Уральцева Н.Н.* Линейные и квазилинейные уравнения параболического типа// М.: Наука, (1967).

# СОДЕРЖАНИЕ